# Computationally Designed Zirconium Organometallic Catalyst for Direct Epoxidation of Alkenes without Allylic H Atoms: Aromatic Linkage Eliminates Formation of Inert Octahedral Complexes

Bo Yang and Thomas A. Manz*

Department of Chemical & Materials Engineering, New Mexico State University, Las Cruces, NM 88003-8001.

*corresponding author email: tmanz@nmsu.edu, phone: (575) 646-2503, fax: (575) 646-7706

---

## Abstract

We used density functional theory to computationally design a Zr organometallic catalyst for selectively oxidizing substrates using molecular oxygen as oxidant without coreductant. Each selective oxidation cycle involves four general steps: (a) a peroxo or weakly adsorbed $O_2$ group releases an O atom to substrate to form substrate oxide and an oxo group, (b) an oxygen molecule adds to the oxo group to generate an $\eta^2$-ozone group, (c) the $\eta^2$-ozone group rearranges to form an $\eta^3$-ozone group, and (d) the $\eta^3$-ozone group releases an O atom to substrate to form substrate oxide and regenerate the peroxo or weakly adsorbed $O_2$ group. This catalyst could potentially be synthesized via the condensation reaction $Zr(N(R)R')_4$ + 2 $C_6H_4$-1,6-(N($C_6H_3$-2',6'-(CH($CH_3$)$_2$)$_2$)OH)$_2$ → Zr($C_6H_4$-1,6-(N($C_6H_3$-2',6'-(CH($CH_3$)$_2$)$_2$)O)$_2$)$_2$ [aka Zr_Benzol catalyst] + 4 N(R)(R')H where R and R' are $CH_3$, $CH_2CH_3$, or other alkyl groups. For direct ethylene epoxidation, the computed enthalpic energetic span (i.e., effective activation energy for the entire catalytic cycle) is 27.1 kcal/mol, which is one of the lowest values for catalysts studied to date. We study reaction mechanisms and the stability of different catalyst forms as a function of the oxygen atom chemical potential. Notably, an aromatic linkage in each ligand prevents this catalyst from deactivating to form an inactive octahedral-like structure that contains the same atoms as the dioxo complex, Zr(Ligand)$_2$(O)$_2$. Due to a side reaction that can transfer an allylic H atom from alkene to catalyst, this catalyst is useful for directly epoxidizing alkenes such as ethylene that do not contain allylic H atoms. To better understand the reaction chemistry, we computed net atomic charges and bond orders for the two catalytically relevant reaction cycles. These results quantify electron transfer and bond forming and breaking during the catalytic process.

---

**Acknowledgements**: Supercomputing resources were provided by the Extreme Science and Engineering Discovery Environment (XSEDE). XSEDE is funded by NSF grant OCI-1053575. XSEDE project grant TG-CTS100027 provided allocations on the Stampede cluster at the Texas Advanced Computing Center (TACC) and the Trestles and Comet clusters at the San Diego Supercomputing Center (SDSC). The authors sincerely thank the technical support staff of XSEDE, TACC, and SDSC. The authors also thank Dr. Karen Goldberg and Wilson Bailey for useful discussions regarding the proposed catalyst synthesis reaction.

**Online Resource 1**: DFT-optimized geometries and energies; imaginary frequency for each transition state; triplet-quintet crossing curves for $O_2$ addition to the oxo complex; table of computed relative energies ($E_{SCF}$, $E_{ZP}$, H, and G) for the Zr_Benzol catalyst with various oxygen-comprised adsorbates; table of assigned spin magnetic moments for triplet complexes; junior and master catalytic cycles and relative energy profiles for direct propene epoxidation using the Zr_Benzol catalyst.

**Online Resource 2**: A 7z format archive containing .xyz files (which can be read using any text editor or the free Jmol visualization program downloadable from jmol.sourceforge.net) containing net atomic charges, bond orders, and atomic spin moments (for spin-polarized systems) for all of the DFT-optimized geometries.

## 1 Introduction

Catalytic selective oxidations is one of the most important reaction types in the chemical industry. [1-3] In spite of the long history of catalytic selective oxidations, key challenges still remain. [1-3] One problem that people have been working on for many years is the utilization of oxidizing agents that are more economical and environmentally friendly. Direct selective oxidation using molecular oxygen as the oxidant without requiring a coreductant is desirable for two reasons. First, ambient air or purified molecular oxygen extracted from air are relatively cheap oxidants compared to other commonly used oxidants such as hydroperoxides, and this is economically important for large-quantity mass-produced chemical intermediates. Second, using an $O_2$ molecule without a coreductant does not require forming a coproduct. The distinction between a coproduct and a byproduct is that a coproduct is produced in the same stoichiometric reaction as the desired product whereas a byproduct is produced in a different stochiometric reaction (i.e., a side reaction). Some selective oxidation catalysts can utilize molecular oxygen as the oxidant without a coreductant. [4-12]

Epoxidation of alkenes is of great importance in the chemical industry since epoxides are widely used intermediates for organic synthesis. [13-17] According to the International Energy Agency’s Technology Roadmap, ethylene oxide (EO) and propylene oxide (PO) are among the top 18 large-volume chemicals worldwide in 2010 and should be targeted for energy and greenhouse gas reductions. [18] Ethylene oxide is commercially manufactured by the direct epoxidation of ethylene over a silver-based catalyst using molecular oxygen as oxidant without a coreductant. [12, 19] In contrast, propylene oxide is commercially manufactured in processes that generate a coproduct and require a coreductant or oxidant besides molecular $O_2$. [14, 20]

In two previous articles, several Zr/Hf-based organometallic complexes with oxygen transfer capability were computationally tested for catalyzing direct ethylene and propene epoxidations. [21, 22] These complexes have a Zr or Hf atom that serves as an oxygen transfer site plus two bidentate ligands coordinated to the metal center via N or O atoms. Our previous computations investigated the following ligands (with Ar = $–C_6H_3\text{-}2,6\text{-}{}^iPr_2$): (a) the diimine ligand N(Ar)-CH-CH-N(Ar) aka NCCN, (b) the imine-nitrone ligand N(Ar)-CH-CH-N(Ar)-O aka NCCNO, and (c) the dinitrone ligand O-N(Ar)-CH-CH-N(Ar)-O aka ONCCNO. [21, 22] The notation Zr/Hf_LTYPE denotes a catalyst system containing a Zr/Hf metal atom bound to two LTYPE ligands (e.g., Zr/Hf_NCCN, Zr/Hf_NCCNO, Zr/Hf_ONCCNO). The Zr_NCCN system was originally proposed and synthesized by Stanciu et al. [23] Unfortunately, it has activation barriers so high that it is inactive for selective oxidation reactions. [21, 22] Lubben and Wolczanski reported molecular oxygen activation by Zr/Hf organometallic complexes with subsequent insertion of one oxygen atom into a tethered olefin group to form an epoxide that remained covalently bound to the metal and insertion of the other oxygen atom into a ligand methyl group to form a methoxy group, but this is a stoichiometric rather than a catalytic process. [24,25] Our computations showed direct ethylene epoxidation barriers for the Zr/Hf_NCCNO systems are substantially lower than for Zr_NCCN,[21, 22] but synthesis of Zr/Hf_NCCNO may be difficult due to coordination of the bidentate ligand via one N and one O atom. It might be easier to synthesize complexes in which each bidentate ligand coordinates via two N atoms (e.g., Zr/Hf_NCCN ligand) or via two O atoms (e.g., Zr/Hf_ONCCNO). However, our calculations showed the Zr/Hf_ONCCNO systems form an inert octahedral complex (containing the same atoms as the dioxo complex) that inhibits the desired selective oxidation reactions. [21, 22] Therefore, in this work, we propose and study a Zr-based catalyst with new ligand architecture that has the following key properties: (a) the bidentate ligand coordinates to the Zr atom via two O atoms, (b) the system does not form an octahedral complex containing the same atoms as the dioxo complex, and (c) there is a straightforward reaction to potentially synthesize the catalyst. This catalyst, $Zr(C_6H_4\text{-}1,6\text{-}(N(C_6H_3\text{-}2',6'\text{-}(CH(CH_3)_2)_2)O)_2)_2$ (aka Zr_Benzol), has an aromatic ring in the ligand backbone that links the two N(O)Ar groups in each ligand. The oxo complex of this catalyst is illustrated in Fig. 1. As in our previous work, we use density functional theory (DFT) to study reaction mechanisms and energetics for direct ethylene and propene epoxidations using molecular oxygen as oxidant without coreductant. Ethylene and propene are the two simplest molecules in the olefin group, which keeps the computational cost to a minimum. Moreover, these represent important examples of alkenes with (i.e., propene) and without (i.e., ethylene) allylic hydrogen atoms.

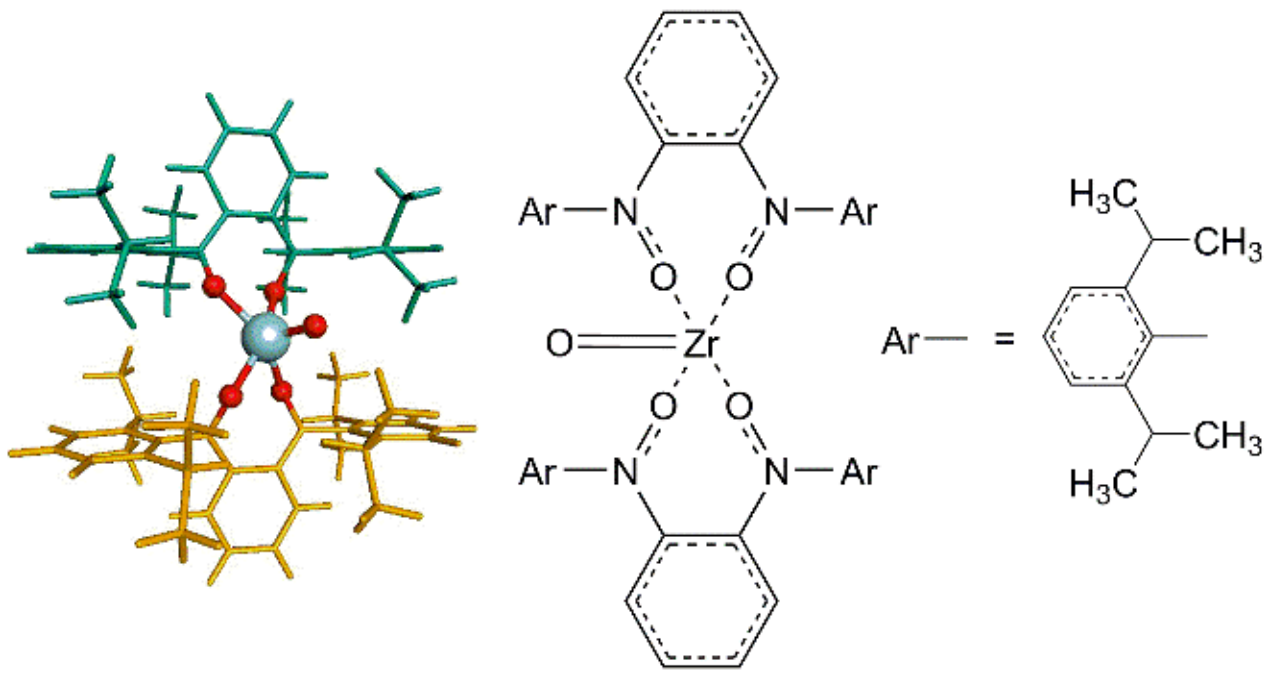

**Fig. 1** The oxo form of the Zr_Benzol catalyst. The triplet oxo complex ($MO^T$) is the catalyst resting state over the majority of the oxygen chemical potential range. The Zr metal atom is colored cyan. The oxygen atoms coordinated to Zr are colored red. All other portions of the ligands are colored green and yellow

## 2 Methods

All calculations were carried out using DFT calculations in GAUSSIAN software. [26] Becke's three-parameter hybrid method involving the Lee-Yang-Parr correlation functional [27, 28] (B3LYP) and LANL2DZ basis sets [26] were chosen to achieve a good balance between high geometry accuracy and low computational cost. Even so, the computations described in this paper took >200000 computer processor hours to complete. Repeating all the calculations with a substantially larger basis set would be extremely computationally expensive. Except where indicated otherwise, catalyst forms were studied in vacuum. Geometries and energies of ground states, transition states, and other reactive intermediates were computed to study reaction mechanisms and activation barriers.

The DFT calculation for each ground and transition state was conducted as described previously. [21] Briefly, for ground state calculations, various initial geometries were considered and full geometry optimizations were carried out to determine the lowest energy state. For each transition state, constrained optimizations were initially performed to generate a geometry estimate that was subsequently optimized using the quadratic synchronous transit or eigenmode following methods. Geometries were optimized in vacuum to better than 0.005 Å for the atom displacements and 0.0025 a.u. for the forces. However, in rare cases with extremely difficult convergence, the geometry was considered converged when the root-mean-squared (rms) force was $< 10^{-4}$ a.u.

Frequency analysis was performed on each ground state and transition state. We verified that all frequencies are positive for each ground state and only one imaginary frequency exists (within a computational tolerance of 30 $cm^{-1}$) for each transition state. Thermochemical analysis was performed under standard condition (pressure of 1 atm and temperature of 298.15 K ) using the harmonic approximation (as implemented in GAUSSIAN 09). The electronic energy without zero-point or thermal vibrational corrections ($E_{SCF}$), the electronic energy including zero-point vibrational correction ($E_{zp}$), the enthalpy (H), and the Gibbs free energy (G) for each optimized geometry are reported in the Online Resource 1.

For each transition state, the imaginary frequency was animated in GaussView to verify its vibration was along the desired reaction pathway. Also, intrinsic reaction coordinate (IRC) calculations were attempted for each of the transition states in all of the ethylene catalytic cycles using GAUSSIAN 09. Specifically, we were able to converge the IRC calculations for transition states $TS_1$, $TS_3$, $TS_6$, $TS_8$, $TS_9$, $TS_{11}$, $TS_{13}$, $TS_{14}$, $TS_{16}$, $TS_{17}$, $TS_{18}$, $TS_{19}$, $TS_{20}$ and $TS_{22}$ as presented in Figs. 7 to 11 using the HPC algorithm [29, 30]. For other transition states, we did try other IRC options in GAUSSIAN 09, but owing to the large size of our system, IRC calculations for other transition states were too computationally expensive and tricky to be converged. For those cases where we were unable to converge IRC calculations, we manually displaced each transition state towards both sides of the reaction pathway along its imaginary vibrational mode, and conducted a full geometry optimization to let it converge to its local energy minimum. In all cases, these IRC calculations or geometry optimizations converged back to the reactants and products, thereby confirming our reaction mechanisms and the correct transition state structure.

As explained in prior literature, for reactions in which the reactants and products have different spin states (aka ‘two-state reactivity’ [31]), the potential energy surfaces defined for these different spin states will cross along the reaction pathway. If this crossing occurs at a relatively low energy along the reaction pathway, the reaction barrier will be characterized by a regular transition state for one of the spin states. On the other hand, if this crossing occurs at a relatively high energy along the reaction pathway, the reaction barrier will be characterized by a minimum energy crossing point between the two spin states. [32, 33]

As indicated by constrained geometry optimization (see Fig. S1, Online Resource 1), there is no regular transition state for $O_2$ addition to the triplet oxo ($MO^T$) complex. In this case, the activation barrier was defined by the triplet-quintet crossing point where a vertical transition (i.e., at constant geometry) occurred. As shown in Fig. S1 (Online Resource 1), this crossing point was estimated by a series of constrained geometry optimizations over the potential energy surface. Starting with the fully optimized $M(\eta^2\text{-}O_3)^T$ structure, two distances were constrained simultaneously: distance "a": the distance between the outer O atom (the one that remains adsorbed to form the oxo group) and the middle O atom, and distance "b": the distance between the Zr atom and the other outer O atom (the one that desorbs to form an $O_2$ molecule). All other geometric parameters were relaxed for both triplet and quintet constrained geometry optimizations. Accordingly, computed SCF energy ($E_{SCF}$) for each constrained geometry was plotted against independently constrained values of the two variables to create a two-dimensional potential energy surface. The triplet energies for constrained optimized triplet and quintet geometries, and quintet energies for constrained optimized triplet and quintet geometries were computed. As shown in Fig. S1 (Online Resource 1), the crossing point was estimated as the lowest energy for which triplet and quintet energies were the same for the same geometry along this potential energy surface.

## 3 Results and Discussion

### 3.1 Key Intermediates and Chemical Potential Diagram

Several key points must be fulfilled by the catalyst in order to perform direct ethylene epoxidation by using only ethylene and $O_2$ as reactants. First, the catalyst should be able to react with $O_2$ molecules to form oxygenated complexes. This should produce a sufficiently weakened bond between two adjacent oxygen atoms so an ethylene molecule could easily extract one of the oxygen atoms from the complex without paying a huge energy penalty for breaking the O-O bond. After giving out the first oxygen atom, the catalyst must also be able to transfer a second oxygen atom (since there are two oxygen atoms in an $O_2$ molecule) to another ethylene molecule with a low energy barrier. In this section, we introduce oxygenated intermediates within our Zr_Benzol catalytic system that possess these properties.

Several conformations of bisperoxo and peroxo-ozone complexes are illustrated in Fig. 2. Three different bisperoxo conformations are achieved: spiro ($M(O_2)_{2(spiro)}$), planar ($M(O_2)_{2(planar)}$), and butterfly ($M(O_2)_{2(butterfly)}$). In the $M(O_2)_{2(spiro)}$ complex, the two peroxo groups bond on opposite sides of the Zr metal with an angle of ~60° between the two O-O bonds. In the $M(O_2)_{2(planar)}$ complex, this angle is ~0° indicating the two peroxo groups and the Zr metal atom are almost coplanar. In the butterfly conformation, the two O-O bonds in the peroxo groups are nearly parallel to each other, but the two bidentate linkages attach to the same side of the Zr metal atom causing the peroxo groups to be lifted up like the wings of a butterfly.

Two ozone conformations ($\eta^2$- and $\eta^3$-ozone) exist within our catalytic system. These ozone groups could potentially pair with one oxygen atom (to form oxo ozone complexes), two oxygen atoms (to form peroxo (or weakly adsorbed $O_2$) ozone complexes) or with nothing (to form ozone complexes) on the opposite side of the Zr metal. In the $\eta^2$-ozone group, only the outer two oxygen atoms are bound to the Zr metal atom and the middle oxygen atom (μ-O) is not bound to Zr. For the $\eta^2$-ozone group, the dihedral angle between the O-O-O plane and the O-Zr-O plane is >135°. With a small energy change, an $\eta^2$-ozone group can transform to generate an $\eta^3$-ozone group. In the $\eta^3$-ozone group, all three oxygen atoms are bound to the Zr metal atom. For the $\eta^3$-ozone group, the dihedral angle between the O-O-O plane and the O-Zr-O plane is <135°. Fig. 2 (d) and (e) shows peroxo $\eta^2$-ozone and peroxo $\eta^3$-ozone structures, respectively. In the peroxo $\eta^3$-ozone complex, the three oxygen atoms in the $\eta^3$-ozone group are already in the position that when one of the outer oxygen atoms is removed by ethylene, the middle oxygen atom along with the other remaining oxygen atom would form the spiro or planar bisperoxo complex depending on which of the two outer oxygen atoms was removed.

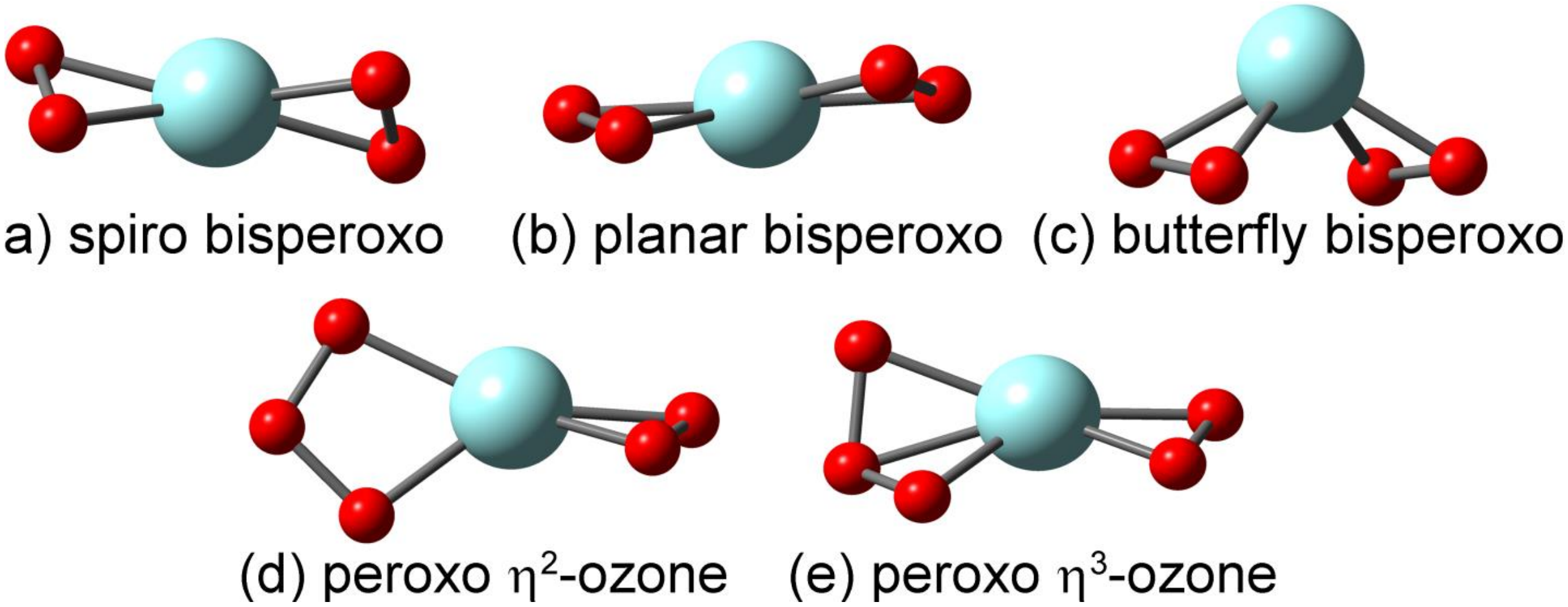

**Fig. 2** Three different bisperoxo conformations of the Zr_Benzol catalyst: (a) spiro, (b) planar, and (c) butterfly. (light blue: Zr, red: O) Two different peroxo ozone conformations of the Zr_Benzol catalyst: (d) peroxo $\eta^2$-ozone and (e) peroxo $\eta^3$-ozone. Ligands are included in the calculations but are not presented here for display purposes

As shown by constrained geometry optimizations, a singlet oxo $\eta^3$-ozone ground state does not exist within this catalytic system. Specifically, the distance between the Zr atom and the μ-O atom (the center oxygen atom in the ozone group) was frozen at a series of different lengths.

All of the other geometric parameters were fully relaxed to minimize the energy. The energy is plotted against the constrained distance in Fig. 3. As we can see in Fig. 3, only one minimum (which represents the oxo η$^2$-ozone conformation) was achieved and no other ground state or transition state was observed. We also performed a search for a triplet oxo η$^3$-ozone structure. Starting with an initial guess for a triplet oxo η$^3$-ozone structure, the geometry optimization converged instead to a triplet oxo η$^2$-ozone complex ($M(O)(\eta^2\text{-}O_3)^T$), which indicates a triplet oxo η$^3$-ozone complex may not exist for this catalyst.

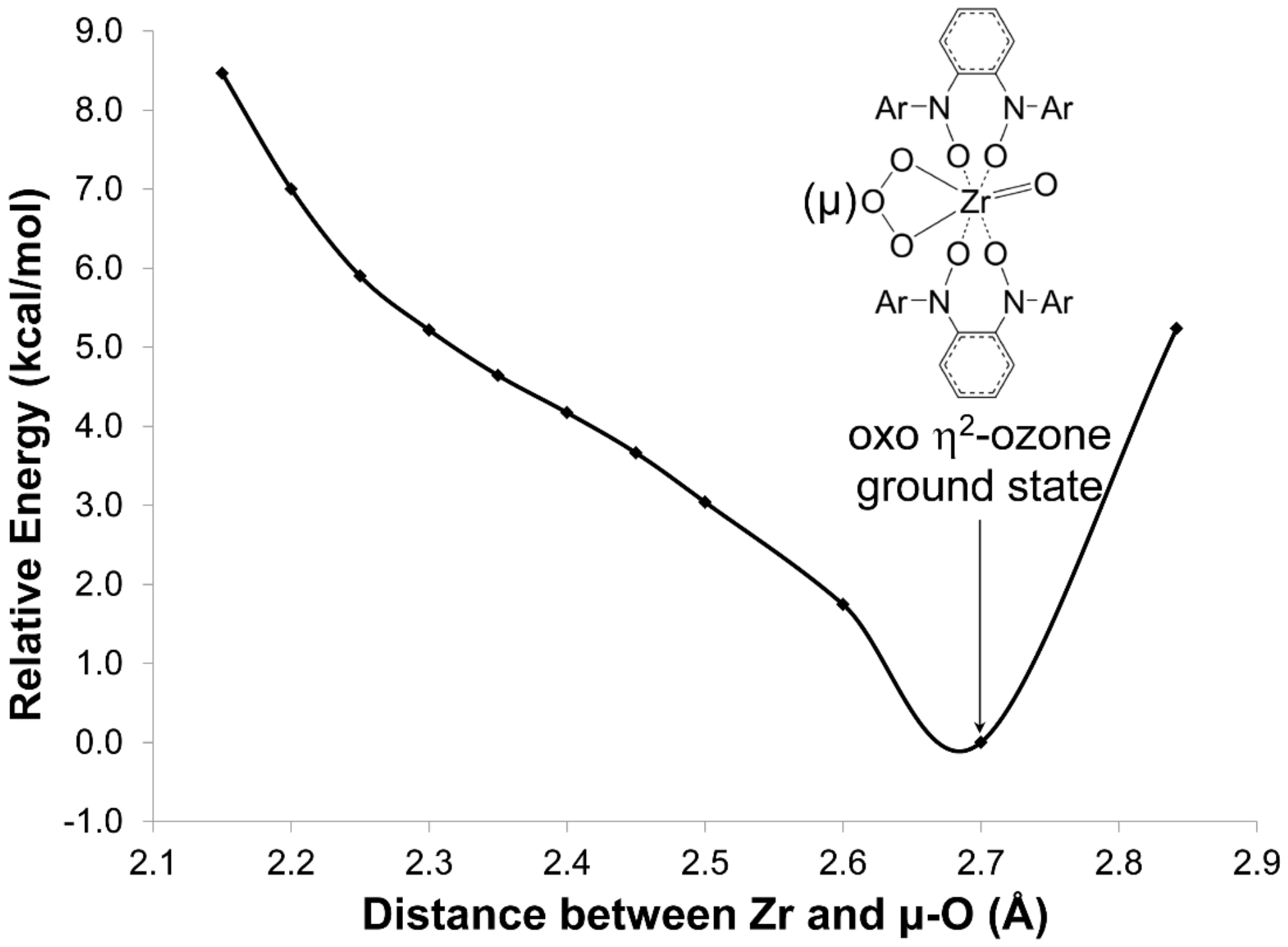

**Fig. 3** Constrained geometry optimization that shows the singlet oxo η$^3$-ozone complex does not exist within the Zr_Benzol system

A chemical potential diagram (Fig. 4) was computed to study the relative energies of different catalyst forms as a function of the oxygen atom chemical potential. The x-value represents the oxygen atom chemical potential. The chemical potential of an oxygen atom in an $O_2$ molecule was chosen as the reference state of 0 kcal/mol and referred to as the $O_2$ side (aka $O_2$-rich side). Accordingly, the oxygen atom chemical potential for the EO side (aka $O_2$-starved side) is the energy of the reaction ethylene + ½ $O_2$ → EO which is -13.8 ($E_{SCF}$), -11.3 ($E_{ZP}$), -12.2 (H), and -6.2 (G) kcal/mol. The y-value represents the energy for each intermediate relative to the singlet spiro bisperoxo complex. On the $O_2$-rich side, the relative energy for each structure (A) was calculated according to the formula:

$$E_A - \frac{N}{2} \cdot E_{O_2} - E_{bisperoxo}.$$

On the $O_2$-starved side, the relative energy for each structure (A) was calculated according to the formula:

$$E_A - N \cdot (E_{ethylene\ oxide} - E_{ethylene}) - E_{bisperoxo}.$$

Here, N represents the difference of oxygen atom number between structure (A) and the bisperoxo complex. E is the DFT-computed energy calculated for the corresponding intermediate. Fig. 4 displays results based on $E_{SCF}$. Table S1 (Online Resource 1) displays the corresponding information based on $E_{SCF}$, $E_{ZP}$, H, and G. On the $O_2$-rich side, complexes with higher relative energies tend to form complexes with lower relative energies by releasing or reacting with $O_2$ molecules. On the $O_2$-starved side, complexes with higher relative energies tend to react with ethylene to produce ethylene oxide plus complexes with lower relative energies.

For convenience, we use M to represent the Zr metal complex with bis(bidentate) ligands. (O) represents an oxo group. ($O_2$) represents a peroxo or weakly adsorbed $O_2$ group. ($\eta^2$-$O_3$) and ($\eta^3$-$O_3$) represent an $\eta^2$- and $\eta^3$-ozone group, respectively. Subscript numbers show the quantity of the corresponding groups. The symbol "·" is placed in front of a weakly adsorbed group. Text in parentheses indicate different geometric conformations. For example, $M{\cdot}(O_2)_{2(spiro)}$ represents a Zr_Benzol intermediate with two weakly adsorbed $O_2$ groups arranged in a spiro conformation. In Fig. 4, singlet states are displayed in the left panel and triplet states are displayed in the middle panel. The right panel is a partial copy of the left panel to enable direct comparison of singlet and triplet energies. In the remainder of this article, we use superscripts S and T to denote singlet and triplet states, respectively.

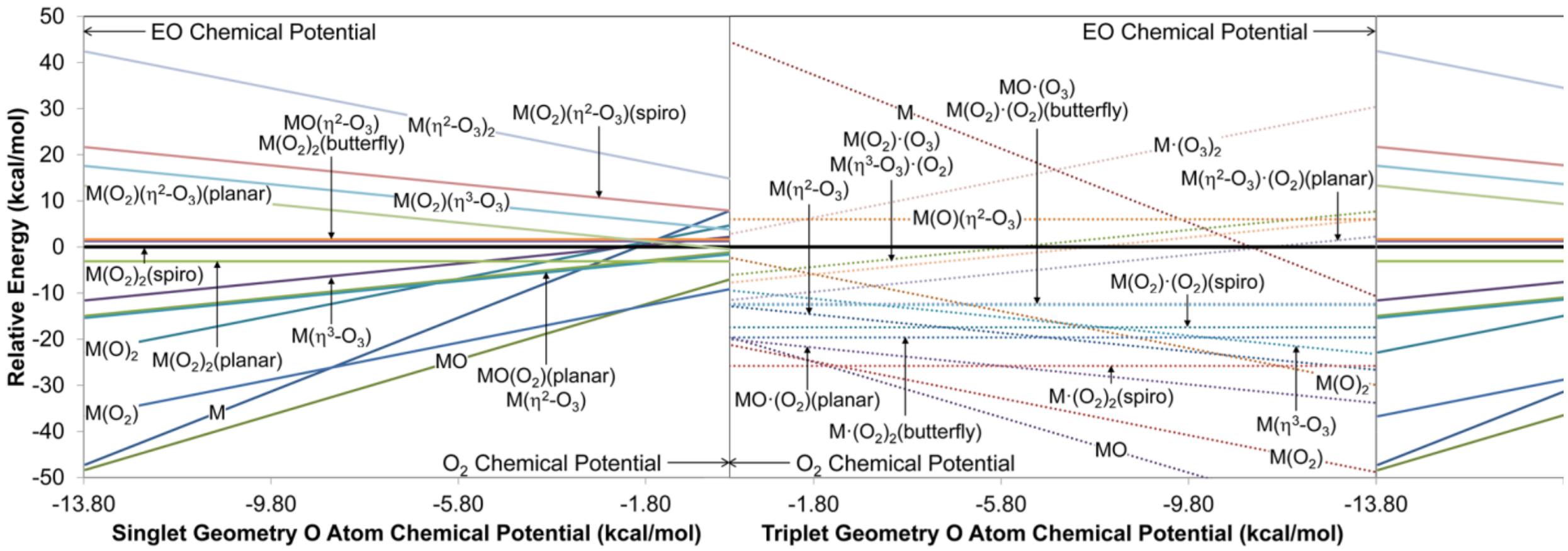

**Fig. 4** Computed chemical potential diagram for Zr_Benzol catalyst. Singlet and triplet forms are displayed in the left and center panel, respectively. The right panel is a partial copy of the left panel for easy comparison. The singlet spiro bisperoxo complex ($M(O_2)_{2(spiro)}$) is the reference state

As shown in Fig. 4, the triplet oxo complex ($MO^T$) is the preferred structure among all catalyst forms across the majority of oxygen chemical potential range. Among the singlet states, the singlet oxo complex ($MO^S$) holds the lowest energy across the majority of oxygen chemical potential range, but its energy is substantially higher than the triplet oxo complex. Among the singlet states, the bis-$\eta^2$-ozone complex ($M(\eta^2\text{-}O_3)_2{}^S$) holds a very high energy on both the $O_2$-rich and $O_2$-starved sides, indicating it will not be stable within the catalytic system. It would generate lower energy complexes by either ejecting $O_2$ molecule(s) directly or granting oxygen atom(s) to ethylene to produce ethylene oxide. Among the triple states, the triplet bare complex ($M^T$) possesses the highest relative energy (44.5 kcal/mol) on the $O_2$-rich side. By reacting with $O_2$ molecule(s), $M^T$ would easily form oxygenated complexes. Among the triplet states, the complex with two weakly adsorbed $\eta^2$-ozone groups ($M{\cdot}(O_3)_2{}^T$) possesses the highest relative energy (34.6 kcal/mol) on the $O_2$-starved side. Major singlet and triplet intermediates of the catalyst with adsorbed oxygenates containing one to six oxygen atoms settle within a ~35 kcal/mol window on the $O_2$-rich end. This suggests the Zr_Benzol system should work efficiently as an oxygen transfer catalyst.

Spin magnetic moments of the triplet states are listed in Table S2 (Online Resource 1) to quantify the distribution of spin magnetization for the following groups of atoms: (a) Zr metal atom, (b) strongly adsorbed O groups, (c) weakly adsorbed O groups, (d) N atoms in ligand 1, (e) N atoms in ligand 2, and (f) the remainder of the structure. For each of these atom groups, the spin magnetic moment was computed by summing the atomic spin moments (ASMs) for all atoms in the group. ASMs were computed using the Density Derived Electrostatic and Chemical (DDEC6) method.[34-37] In the table, ligand 1 is arbitrarily designated as the ligand having larger ASMs for nitrogen atoms. The combined ASMs for all parts sum to 2.00, representing the two unpaired electrons. As shown in Table S2 (Online Resource 1), except for the bare complex ($M^T$), the Zr atoms in other structures have ASM magnitudes ≤ 0.07. In general, a large portion of the spin density resides on weakly adsorbed non-ligand oxygen atoms. Ligand nitrogen atoms are the second dominant parts for occupying the spin. In all of the structures except $M(O)^T_2$ and $MO(\eta^2\text{-}O_3)^T$, the strongly adsorbed O atoms hold little spin (ASM magnitudes < 0.05).

**3.2 Investigation of Catalyst Stability**

Only one deformation product, as shown in Fig. 5, was found during our calculations. Normally, the reaction of an oxygen molecule with an oxo group is expected to produce an $\eta^2$-ozone group. However, the $O_2$ molecule can also attack the oxo group in such a way that one O atom from $O_2$ bonds with the oxo group and the other O atom attacks the ligand C–C bond (instead of the Zr metal). This forms a heterocyclic ring containing the metal center and part of the ligand as highlighted in red in Fig. 5. Because the energy of this cyclized structure is 2 kcal/mol higher than the corresponding $\eta^2$-ozone complex, this ligand cyclization product would be unstable with respect to other catalyst forms. Therefore, this ligand cyclization reaction is energetically unimportant.

In previous reports, we showed that Zr/Hf_ONCCNO complexes containing bis(dinitrone) ligation form inert octahedral complexes that have energies >20 kcal/mol lower than all intermediates involved in the main catalytic cycles. [21, 22] Because these inert octahedral complexes must transform back into the dioxo complexes or other intermediates in the catalytic cycle before epoxidation can occur, their existence causes the overall energetic spans for the corresponding catalysts to be ~20 kcal/mol plus the cycle activation barrier. This makes the Zr/Hf_ONCCNO catalysts unsuitable for alkene epoxidation.

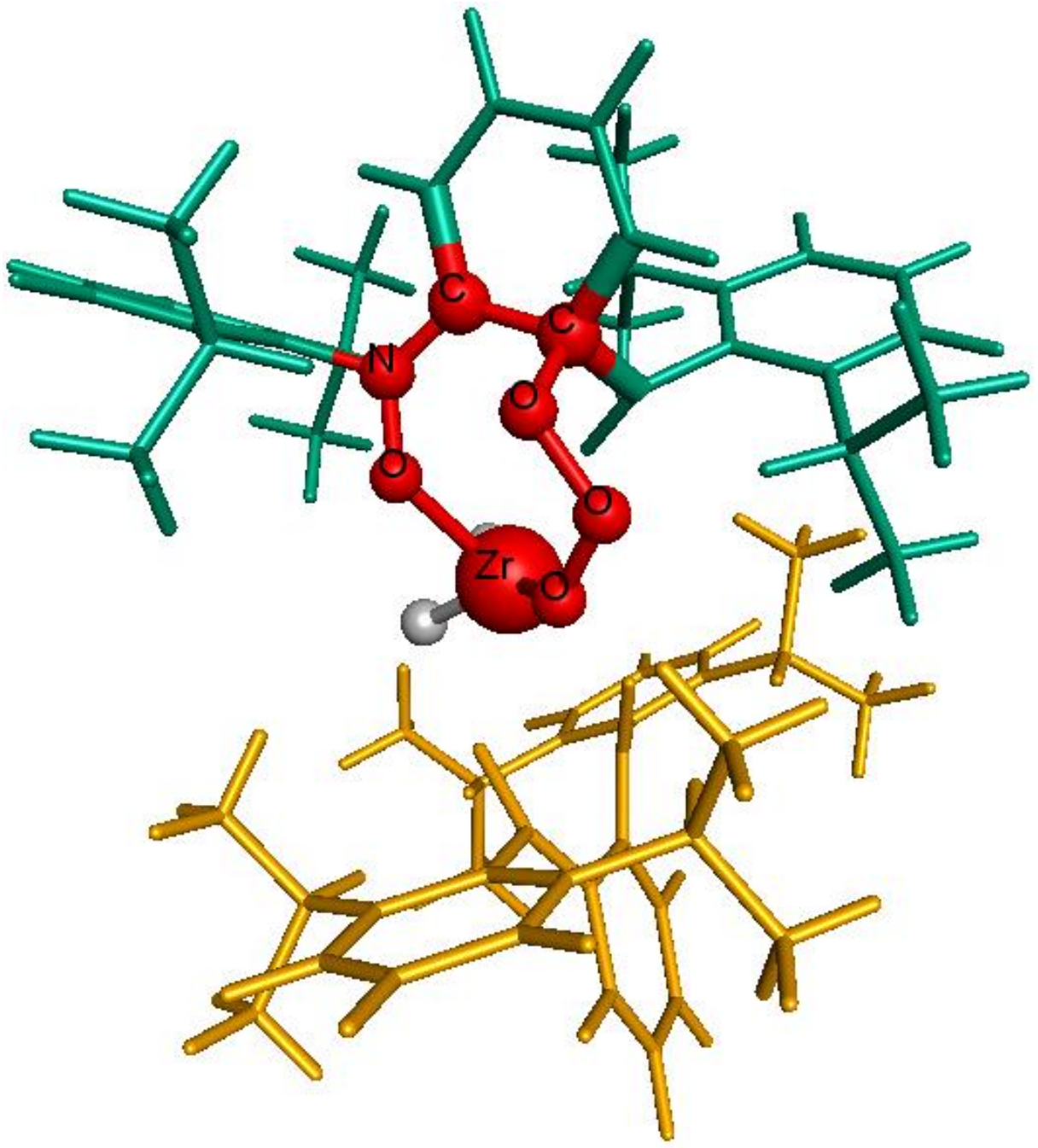

**Fig. 5** Optimized structure for ligand cyclization product with heterocyclic ring highlighted in red. Remaining atoms of this ligand are colored green. The other organic ligand is colored yellow. Other oxygen atoms bound to the metal are colored grey

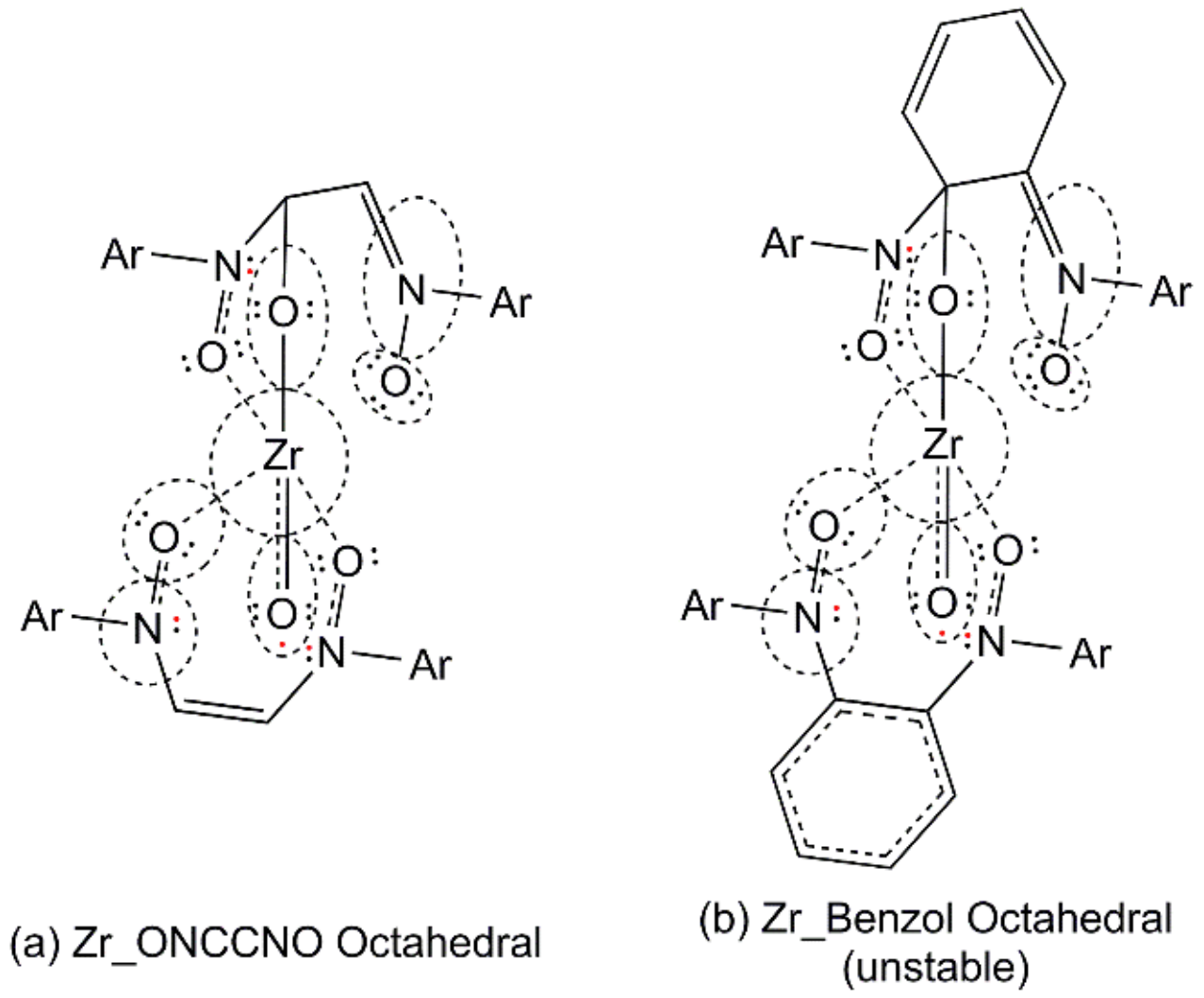

**Fig. 6** Structural cartoons of octahedral conformations for (a) Zr_ONCCNO and (b) Zr_Benzol catalysts. Each pair of black dots represents an electron pair. A black dot paired with one red dot represents 1.5 electrons which formally corresponds to an ASM of ~0.5. A dashed circle around an atom encloses electrons (in the form of lone pairs and bonds) nominally assigned to that atom. A dashed circle passing through the middle of a bond indicates electrons in the bond are equally assigned to two atoms. A dashed bond indicates a half-order bond. The Zr_Benzol octahedral structure is not a stationary point on the potential energy surface and spontaneously transforms into the dioxo conformation during geometry optimization

Fig. 6(a) is a structural cartoon of the Zr_ONCCNO octahedral complex. Electron pairs and bonds have been assigned to each atom to satisfy the octet principle. The location of the unpaired electrons (indicated by red dots) is corroborated by the previously computed atomic spin moments (ASMs). [21, 22] One of the oxygen atoms is bound to the Zr center plus a ligand C atom. This C atom forms four single-like bonds (with adjacent H, N, C and O atoms) to satisfy the octet principle. Fig. 6(b) is a structural cartoon of a hypothetical octahedral-like conformation for the Zr_Benzol system with bonds around the Zr center similar to those in the Zr_ONCCNO octahedral complex. Notably, the oxygen atom bound to the Zr center plus a ligand C atom causes this C atom to form single-like bonds to the two adjacent C atoms in the ligand. This would destroy the aromaticity of the ligand's hexagonal ring by forcing it into a cyclohexadiene-like bonding motif. Because an aromatic benzene-like ring is energetically more favorable than a nonaromatic cyclohexadiene-like ring, this octahedral-like conformation would accordingly raise the energy of the whole structure. In fact, when we started geometry optimizations with this octahedral-like conformation for the Zr_Benzol catalyst, it always spontaneously reverted to the Zr_Benzol dioxo complex. This indicates this octahedral-like conformation is so unstable that it does not even form a stationary point on the potential energy surface. Thus, the Zr_Benzol catalyst does not deactivate to form an octahedral-like structure that contains the same atoms as the dioxo complex, $ML_2(O)_2$.

### 3.3 Catalytic Cycles and Reaction Energy Profiles

Four interrelated junior cycles which go through four different $\eta^3$-ozone intermediates were investigated and are shown in Figs. 7, 8, 9, and 10. (A junior cycle is a specific reaction pathway that results in complete turn-over of the catalyst. The master catalytic cycle is the entire reaction network containing all junior cycles.) In each of these figures, the top panel displays the reaction mechanism, and the bottom panel displays the reaction energy profile. As described in Section 3.1, $MO^T$ is the catalyst resting state across the majority of the oxygen chemical potential range. Therefore, we have included $MO^T$ as an intermediate in all four junior cycles. Fig. 11 illustrates the full reaction network (i.e., master catalytic cycle) with emphasis on reaction steps that link the four junior cycles. Figs. 7, 8, 9, 10, and 11 display $E_{SCF}$ energies for each reaction step. Transition state and net reaction energies including zero-point and thermodynamic vibrational corrections (i.e., $E_{ZP}$, H, G) are summarized in Table 1 along with $E_{SCF}$.

For each junior cycle, the cycle activation barrier is defined as the largest (multistep) energy barrier along the forward direction of the cycle. The catalyst's energetic span is defined as the minimum cycle activation barrier among all junior cycles. As reviewed by Kozuch and Shaik, the energetic span quantifies the apparent energy barrier for the entire catalytic cycle. [38] As we stated previously, "The energetic span is the energy difference between the turn-over-frequency determining intermediate (TDI) and a subsequent turn-over-frequency determining transition state (TDTS). [38] The determination of TDI and TDTS is performed independently for the SCF energy, $E_{ZP}$, enthalpy, and Gibbs free energy. If we imagine the catalytic cycle as a wheel, we can choose any intermediate as a starting and ending point of the cycle. The TDI is that intermediate, which if chosen as a starting point, leads to the highest subsequent transition state energy (i.e., the TDTS) along the preferred catalytic cycle before returning. [38] The preferred catalytic cycle should be chosen to contain the catalyst resting state under reaction conditions."[21]

As shown in Fig. 7, the first junior cycle ($1^{st}$ cycle) involves $MO\cdot(O_2)^T$, $M(\eta^2\text{-}O_3)\cdot(O_2)^T$, $M(\eta^3\text{-}O_3)\cdot(O_2)^T$, $M\cdot(O_2)^T_{2,spiro}$, and $MO^T$ complexes. $MO^T$ adsorbs one $O_2$ molecule to generate $MO\cdot(O_2)^T$ with an activation barrier of 7.9 kcal/mol and a net reaction energy of 7.9 - 8.4 = -0.5 kcal/mol. By reacting with one $O_2$ molecule, $MO\cdot(O_2)^T$ turns into $M(\eta^2\text{-}O_3)\cdot(O_2)^T$ with an activation barrier of 16.9 kcal/mol and a net reaction energy of 16.9 – 8.5 = 8.4 kcal/mol. Subsequently, $M(\eta^2\text{-}O_3)\cdot(O_2)^T$ transforms into $M(\eta^3\text{-}O_3)\cdot(O_2)^T$ with an activation barrier of 4.7 kcal/mol and a net reaction energy of 4.7 – 1.0 = 3.7 kcal/mol. In the next step, $M(\eta^3\text{-}O_3)\cdot(O_2)^T$ reacts with an ethylene molecule to form EO plus $M\cdot(O_2)^T_{2,spiro}$ with an activation barrier of 12.1 kcal/mol. In the final step, $M\cdot(O_2)^T_{2,spiro}$ reacts with an ethylene molecule to form EO and regenerate $MO\cdot(O_2)^T$ to finish one whole catalytic cycle. The activation barrier for the final reaction step is 21.5 kcal/mol. The cycle activation barrier, as shown in the lower panel of Fig. 7, goes from $MO\cdot(O_2)^T$ to $TS_4$ and equals 16.9 - 8.5 + 4.7 - 1.0 +12.1 = 24.4 kcal/mol.

As shown in Fig. 8, the $2^{nd}$ junior cycle ($2^{nd}$ cycle) involves $MO(O_2)^S$, $MO\cdot(O_2)^T$, $M(O_2)(\eta^2\text{-}O_3)^S$, $M(O_2)(\eta^3\text{-}O_3)^S$, $M(O_2)^S_{2,spiro}$, $M(O_2)^S_{2,planar}$, and $MO^T$ complexes. In the first step, $MO(O_2)^S$ reacts with an $O_2$ molecule to produce $M(O_2)(\eta^2\text{-}O_3)^S$ with an activation barrier of 17.4 kcal/mol and a net reaction energy of 17.4 - 16.7 = 0.7 kcal/mol. $M(O_2)(\eta^2\text{-}O_3)^S$ rearranges to form $M(O_2)(\eta^3\text{-}O_3)^S$ with an activation barrier of 5.3 kcal/mol and a net reaction energy of 5.3 - 1.0 = 4.3 kcal/mol. Then, $M(O_2)(\eta^3\text{-}O_3)^S$ reacts with an ethylene molecule to produce an EO molecule plus $M(O_2)^S_{2,spiro}$ with an activation barrier of 14.0 kcal/mol and a net reaction energy of 14.0 - 31.6 = -17.6 kcal/mol. Two reaction pathways could be followed after forming $M(O_2)^S_{2,spiro}$:

1) $M(O_2)^S_{2,spiro}$ directly reacts with an ethylene molecule to generate an EO molecule plus $MO(O_2)^S$ to finish one whole cycle. The activation barrier for this step is 20.8 kcal/mol and the net reaction energy is 20.8 - 35.7 = -14.9 kcal/mol.

2) $M(O_2)^S_{2,spiro}$ transforms into $M(O_2)^S_{2,planar}$ with an energy barrier of 7.1 kcal/mol and a net reaction energy of 7.1 - 10.2 = -3.1 kcal/mol. $M(O_2)^S_{2,planar}$ then reacts with ethylene to generate an EO plus $MO(O_2)^S$ to finish one whole cycle. The activation barrier for this step is 22.9 kcal/mol and the net reaction energy is 22.9 - 34.8 = -11.9 kcal/mol.

$MO^T$ is included as the catalyst resting state. $MO^T$ forms an equilibrium with $MO\cdot(O_2)^T$ with the same mechanism as in the 1st cycle. $MO\cdot(O_2)^T$ transforms into its singlet state ($MO(O_2)^S$) by paying 18.8 kcal/mol of energy to join the 2nd cycle . The cycle activation barrier, as shown in the lower panel of Fig. 8, goes from $MO\cdot(O_2)^T$ to $TS_9$ and equals 18.8 + 17.4 – 16.7 + 5.3 – 1.0 + 14.0 = 37.8 kcal/mol.

As shown in Fig. 9, the third junior cycle (3rd cycle) involves $MO^T$, $M(\eta^2\text{-}O_3)^T$, $M(\eta^3\text{-}O_3)^T$, $M(O_2)^T$, and $M\cdot(O_2)^T_{2,butterfly}$ intermediates. The cycle starts with $MO^T$ reacting with one $O_2$ molecule to form $M(\eta^2\text{-}O_3)^T$. The energy barrier associated with this reaction step, as determined by the triplet-quintet constrained optimization crossing-point (CP) shown in Fig. S1 (Online Resource 1), is 21.4 kcal/mol and the net reaction energy is 21.4 - 14.8 = 6.6 kcal/mol. In the second reaction step, $M(\eta^2\text{-}O_3)^T$ transforms into $M(\eta^3\text{-}O_3)^T$ with an energy barrier of 4.2 kcal/mol and a net reaction energy of 4.2 - 0.8 = 3.4 kcal/mol. Then, $M(\eta^3\text{-}O_3)^T$ reacts with ethylene to form EO plus $M(O_2)^T$ with an energy barrier of 15.5 kcal/mol and a net reaction energy of 15.5 - 41.4 = -25.6 kcal/mol. Then, $M(O_2)^T$ reacts with ethylene to form EO and regenerate $MO^T$ to finish one whole cycle. The activation barrier for this step is 20.8 kcal/mol and the net reaction energy is 20.8 - 32.9 = -12.1 kcal/mol. Moreover, $M\cdot(O_2)^T_{2,butterfly}$ could join the 3rd cycle by ejecting one $O_2$ molecule to produce $M(O_2)^T$ with an associated activation barrier of 12.1 kcal/mol and a net reaction energy of 12.1 - 20.7 = -8.6 kcal/mol. The cycle activation barrier, as shown in the lower panel of Fig. 9, goes from $MO^T$ to $TS_{14}$ and equals 21.4 – 14.8 + 4.2 – 0.8 + 15.5 = 25.6 kcal/mol.

As shown in Fig. 10, the fourth junior cycle (4th cycle) involves $MO^S$, $MO^T$, $M(\eta^2\text{-}O_3)^S$, $M(\eta^2\text{-}O_3)^T$, $M(\eta^3\text{-}O_3)^S$, and $M(O_2)^S$ intermediates. The reaction begins with a quick transformation between two oxo conformations. By giving out 12.5 kcal/mol of energy, $MO^S$ turns into $MO^T$. Subsequently, $MO^T$ reacts with an $O_2$ molecule to form $M(\eta^2\text{-}O_3)^T$ through the same reaction step as in the 3rd cycle. Then, $M(\eta^2\text{-}O_3)^T$ transforms into its singlet conformation ($M(\eta^2\text{-}O_3)^S$) by absorbing 11.2 kcal/mol of energy. $M(\eta^2\text{-}O_3)^S$ transforms into $M(\eta^3\text{-}O_3)^S$ by overcoming an energy barrier of 4.6 kcal/mol with a net reaction energy of 4.6 - 0.8 = 3.8 kcal/mol. $M(\eta^3\text{-}O_3)^S$ then reacts with ethylene to produce $M(O_2)^S$ plus EO with an energy barrier of 16.0 kcal/mol and a net reaction energy of 16.0 - 41.1 = -25.1 kcal/mol. $M(O_2)^S$ then reacts with ethylene to produce EO and regenerate $MO^S$ to finish one whole cycle. The activation barrier for this step is 20.3 kcal/mol and the net reaction energy is 20.3 - 32.0 = -11.7 kcal/mol. The cycle activation barrier, as shown in the lower panel of Fig. 10, goes from $MO^T$ to $TS_{18}$ and equals 21.4 – 14.8 + 11.2 + 4.6 – 0.8 + 16.0 = 37.7 kcal/mol.

Fig. 11 shows the master catalytic cycle summarizing relationships between the four junior cycles. Each junior cycle consumes one $O_2$ molecule and two ethylene molecules to produce two EO molecules through a catalytic mechanism involving an $\eta^3$-ozone group. $MO^T$, as the catalyst resting state over a wide oxygen chemical potential range, serves as the hub connecting all junior cycles. Other pathways also connect these four junior cycles. $M(O)_2^T$ can be generated by the reaction of ethylene with $MO{\cdot}(O_2)^T$ (from 1st cycle) to produce EO. Similarly, $M(O)_2^S$ can be generated by the reaction of ethylene with $MO(O_2)^S$ (from 2nd cycle) to produce EO. $M(O)_2^S$ can subsequently transform into $M(O)_2^T$ which can rearrange to produce $M(O_2)^T$ to join the 3rd cycle. Another pathway also connects cycles 1, 2, and 3. Specifically, $M(O_2)_{2,spiro}^S$ (in 2nd cycle) can transform into $M{\cdot}(O_2)_{2,spiro}^T$ (in 1st cycle) which can eject an $O_2$ molecule to generate $M(O_2)^T$ (in 3rd cycle). As explained above, the 3rd and 4th cycles share the $MO^T$ and $M(\eta^2\text{-}O_3)^T$ intermediates.

We now summarize the calculation of energetic spans for the Zr_Benzol catalyst. As discussed above, the SCF cycle activation barriers are 24.4 (1st cycle), 37.8 (2nd cycle), 25.6 (3rd cycle), and 37.7 (4th cycle). The first cycle is the preferred catalytic cycle, because it has the lowest cycle activation barrier. Accordingly, our calculated SCF energetic span for direct ethylene epoxidation over the Zr_Benzol catalyst is 24.4 kcal/mol with $MO{\cdot}(O_2)^T$ as TDI and $TS_4$ as TDTS. Because the cycle activation barrier for the 3rd cycle is about the same as for the 1st cycle, the 3rd cycle is also kinetically important. The 2nd and 4th cycles are less kinetically important, because they have cycle activation barriers >10 kcal/mol higher than the 1st and 3rd cycles. The other energetic spans (i.e., $E_{zp}$, H, and G) were computed in a similar manner with the TDI and TDTS determined independently for each energy type. Table 2 compares the $E_{SCF}$, $E_{ZP}$, H, and G energetic spans for the Zr_Benzol catalyst to our previous results [22] for the Zr_NCCN, Zr_NCCNO, and Hf_NCCNO catalysts. The Zr_NCCN complex, which was first synthesized by Stanciu et al.,[23] has an energetic span too high for use as a selective oxidation catalyst. [21] The Hf_NCCNO, Zr_NCCNO, and Zr_Benzol catalysts have computed $E_{SCF}$, $E_{ZP}$, and H energetic spans between 24.4 and 30.5 kcal/mol, with the G energetic spans between 45.3 and 52.9 kcal/mol. Although energetic spans for the Zr/Hf_NCCNO and Zr_Benzol catalysts are similar, we believe the Zr_Benzol catalyst will be easier to synthesize than the Zr/Hf_NCCNO catalysts. (A proposed synthesis reaction is described in Section 4.) For this reason, we believe the Zr_Benzol catalyst is an important improvement over the Zr/Hf_NCCNO catalysts.

Table 1 also summarizes the transition state lateness descriptor ($W_{TS}$) for elementary reactions we encountered in our catalytic system. ($W_{TS}$ is not available for the spin state transitions.) Transition state lateness is used to classify if a transition state is reactant-like (early) or product-like (late). [39] Such information is helpful for future catalyst modification by determining the sensitivity of an activation barrier to the reactant and product structures. [39-42] In our study, the Dimensionless Reaction Coordinate Software (DRCS) provided by Manz and Sholl [39] was used to compute the dimensionless reaction coordinate for the transition state ($W_{TS}$). A transition state is said to be early when $W_{TS} < 0.5$, late when $W_{TS} > 0.5$, and equidistant between reactants and products when $W_{TS} = 0.5$. [39] Based on our calculations, all transition states that related to the $\eta^2$-ozone formation have $W_{TS} \leq 0.5$, which indicates early transition states. Under such circumstance, the activation energy of the $\eta^2$-ozone formation reaction is expected to be more sensitive to the reactant (oxo or oxo peroxo complex) structure than to the product ($\eta^2$-ozone or peroxo $\eta^2$-ozone complex) structure. On the other hand, all transition states for $\eta^2$ to $\eta^3$-ozone transformation have $W_{TS} > 0.5$ indicating late transition states. Accordingly, the activation energy of the ozone transformation reaction is expected to be more sensitive to the product structure than to the reactant structure. Transition states corresponding to ethylene oxide formation can be divided into two categories: a) transition states that have two adsorbed non-ligand oxygen groups with computed $W_{TS}$ around 0.5, and b) transition states that have only one adsorbed non-ligand oxygen group with computed $W_{TS} > 0.5$. For the first category, the energy of the transition state is expected to be sensitive to both the reactant and product structures. For the second category, the transition state energy is expected to be more sensitive to the product structure.

This $W_{TS}$ descriptor is also useful for quantifying whether a chemical reaction follows the Hammond-Leffler postulate. [39] According to the Hammond-Leffler postulate, an endothermic reaction is anticipated to have a late transition state, and an exothermic reaction is anticipated to have an early transition state. [72-73] Examining Table 1, 35% of the reactions followed the Hammond-Leffler postulate and 65% did not.

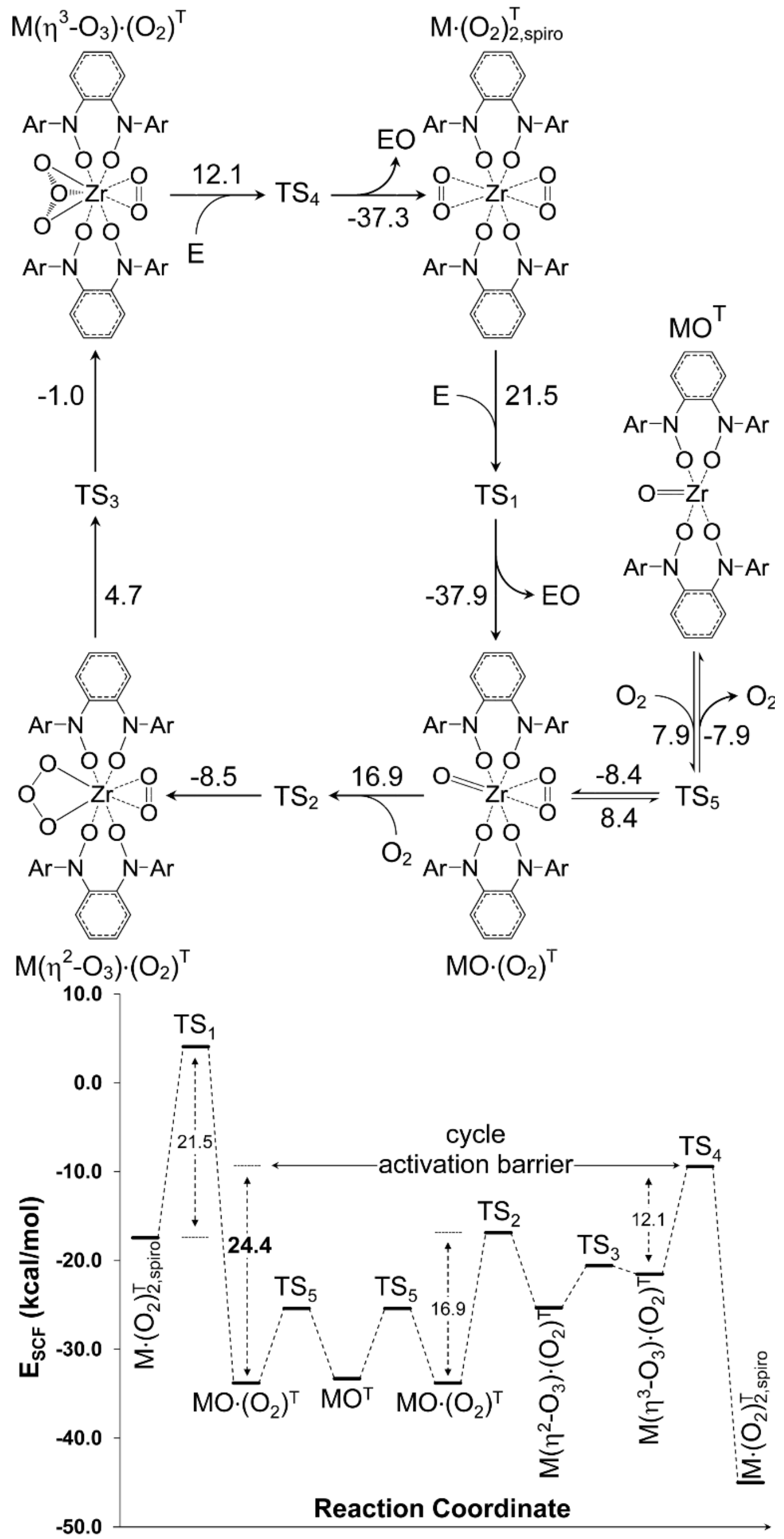

**Fig. 7** (Upper panel) M($\eta^3$-$O_3$)·($O_2$)$^{\mathbf{T}}$ involved junior cycle (1st cycle) for direct ethylene epoxidation. (Lower panel) SCF energy profile for the 1st junior cycle. (In the lower panel, the singlet spiro bisperoxo complex was taken to be the reference state with energy zero.) Energies are presented in kcal/mol

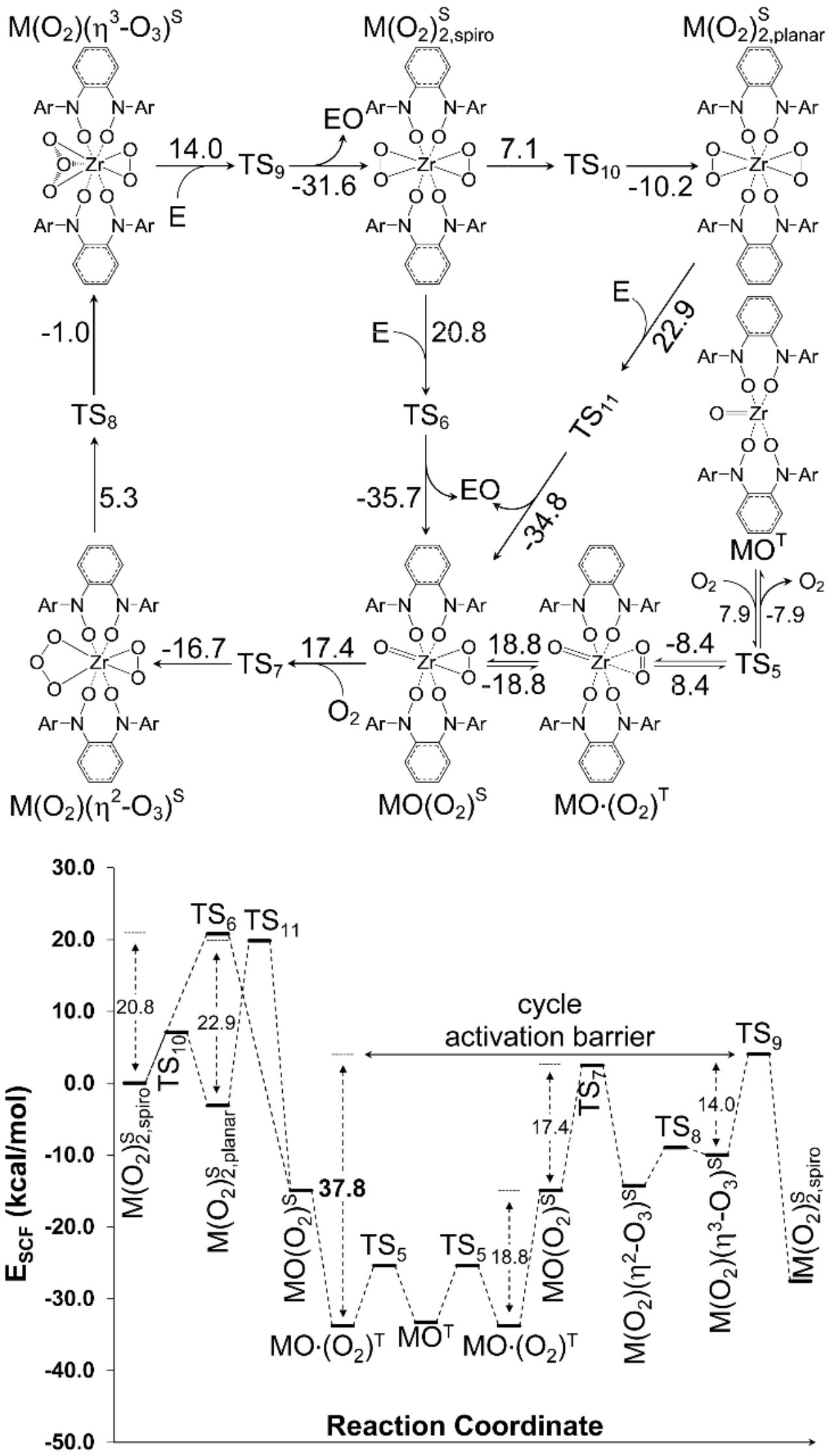

**Fig. 8** (Upper panel) Singlet peroxo η[3]-ozone intermediate ($M(O_2)(\eta^3\text{-}O_3)^S$) involved junior cycle (2nd cycle) for direct ethylene epoxidation. (Lower panel) SCF energy profile for the 2nd junior cycle. Energies are presented in kcal/mol

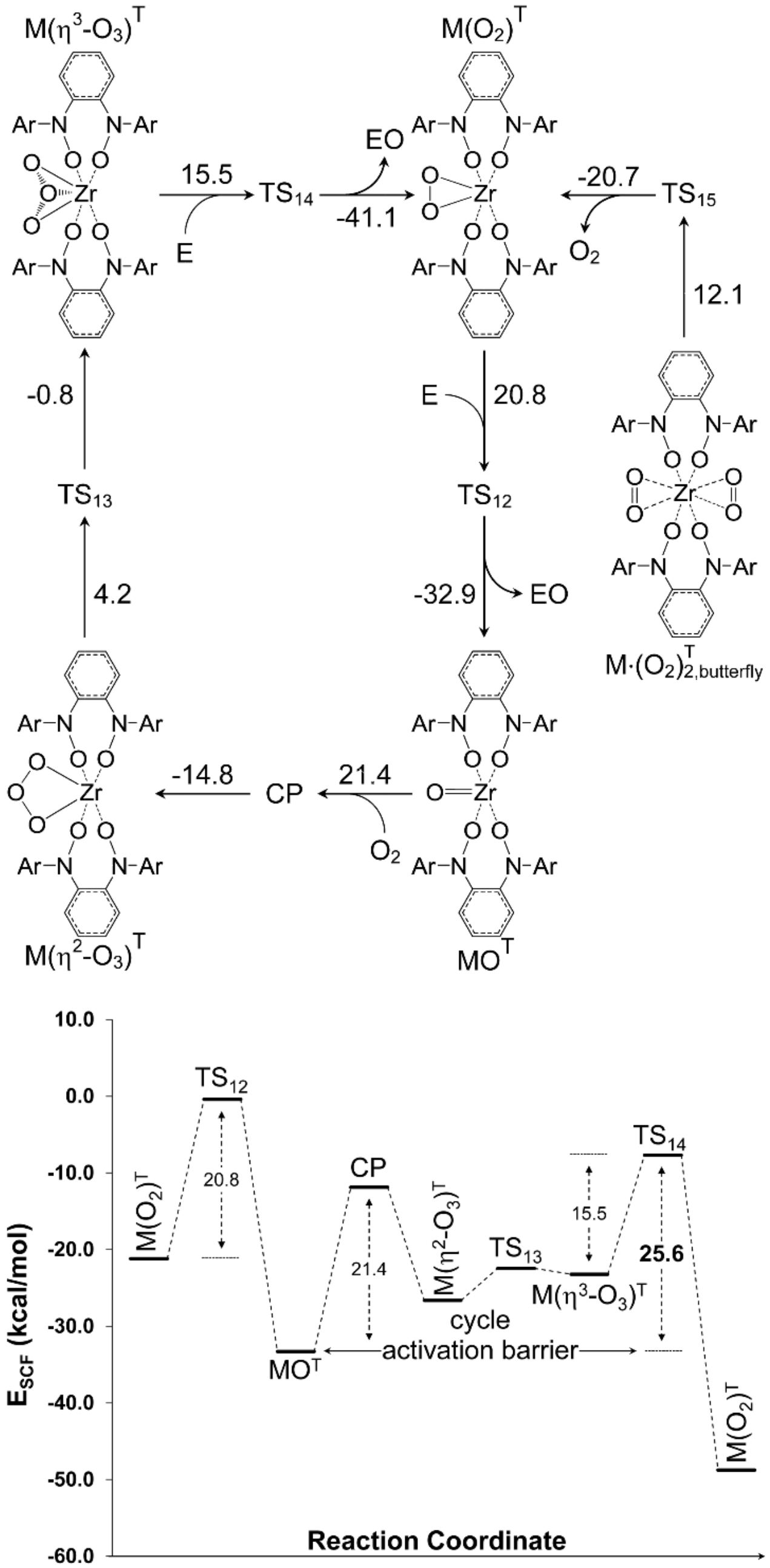

**Fig. 9** (Upper panel) Triplet $\eta^3$-ozone intermediate ($M(\eta^3\text{-}O_3)^T$) involved junior cycle (3rd cycle) for direct ethylene epoxidation. (Lower panel) SCF energy profile for the 3rd junior cycle. (In the lower panel, the singlet spiro bisperoxo complex was taken to be the reference state with energy zero.) Energies are presented in kcal/mol

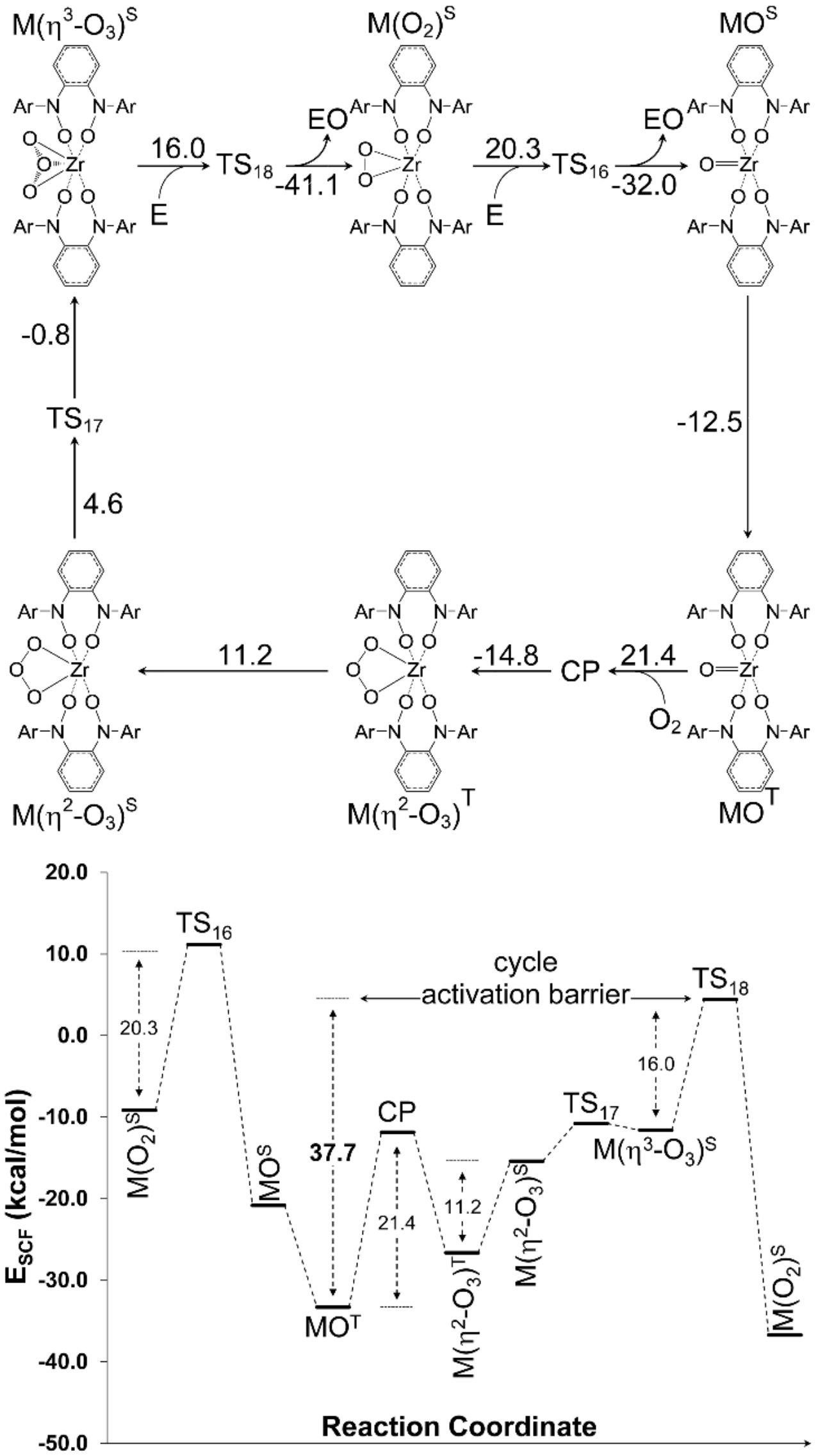

**Fig. 10** (Upper panel) Singlet η$^3$-ozone intermediate (M(η$^3$-$O_3$)$^S$) involved junior cycle (4$^{th}$ cycle) for direct ethylene epoxidation. The triplet oxo (MO$^T$) and η$^2$-ozone (M(η$^2$-$O_3$)$^T$) ground states and the crossing-point (CP) are shared between the 3$^{rd}$ and 4$^{th}$ cycles. (Lower panel) SCF energy profile for the 4$^{th}$ junior cycle. (In the lower panel, the singlet spiro bisperoxo complex was taken to be the reference state with energy zero.) Energies are presented in kcal/mol

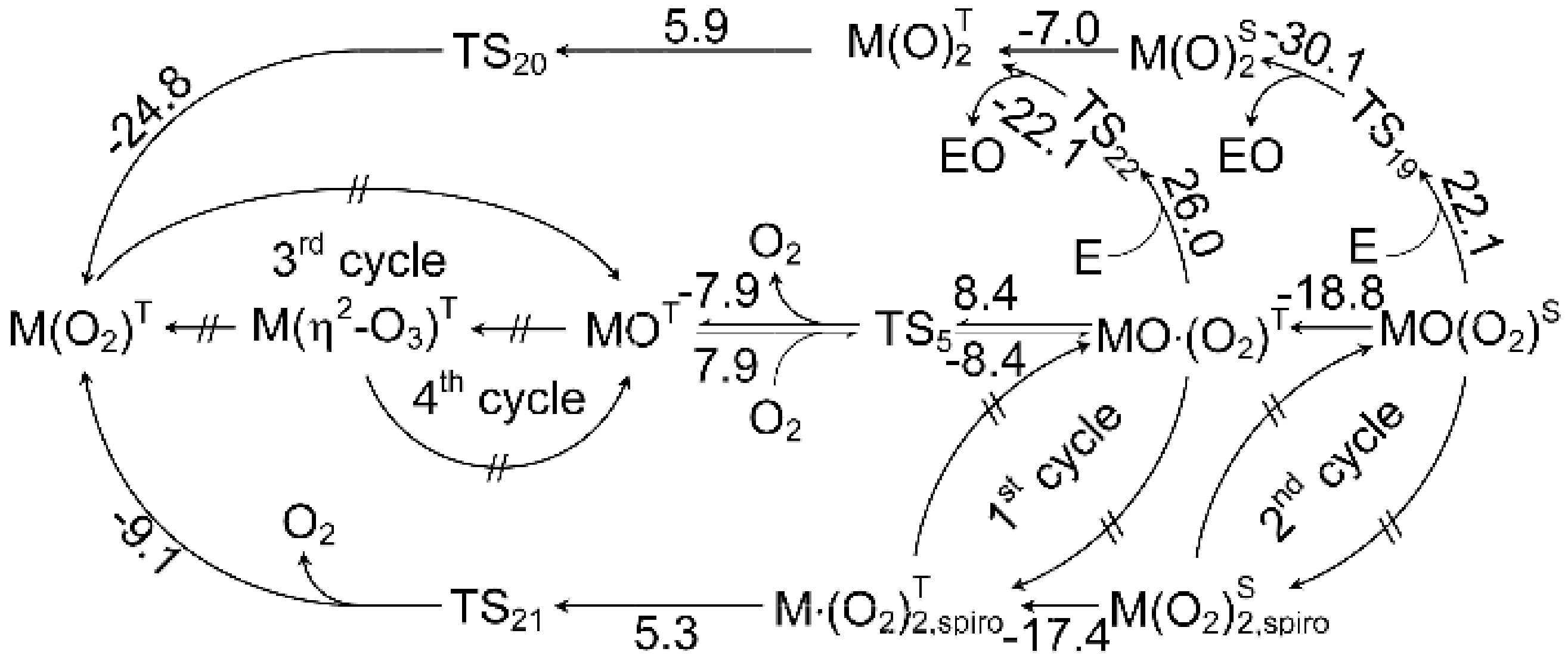

**Fig. 11** Master catalytic cycle for direct ethylene epoxidation using Zr_Benzol catalyst. Energies are presented in kcal/mol

**Table 1** Summary of computed transition state and net reaction energies for various steps in the direct epoxidation of ethylene to EO over the Zr_Benzol catalyst using molecular oxygen as the oxidant. The transition state lateness descriptors ($W_{TS}$) are listed in the last column. For spin state transitions, $W_{TS}$ is not available and only net reaction energies are listed

| reactant | product | activation barrier | | | | net rxn energy | | | | $W_{TS}$ |
|---|---|---|---|---|---|---|---|---|---|---|
| | | $E_{SCF}$ | $E_{ZP}$ | H | G | $E_{SCF}$ | $E_{ZP}$ | H | G | |
| $M(O)\cdot(O_2)^T$ | $M(O)^T+O_2$ | 8.4 | 8.6 | 8.2 | 9.4 | 0.5 | -1.0 | -1.2 | -9.7 | 0.65 |
| $M(O)^T+O_2$ | $M(O)\cdot(O_2)^T$ | 7.9 | 9.6 | 9.4 | 19.1 | -0.5 | 1.0 | 1.2 | 9.7 | 0.35 |
| $M(O)\cdot(O_2)^T+E$ | $M(O)_2^T+EO$ | 26.0 | 27.1 | 26.4 | 39.9 | 3.9 | 8.1 | 7.5 | 8.1 | 0.46 |
| $M(O)_2^T$ | $M(O_2)^T$ | 5.9 | 2.3 | 1.8 | 4.2 | -18.9 | -21.4 | -21.7 | -21.2 | 0.74 |
| $M\cdot(O_2)_{2,spiro}^T$ | $M(O_2)^T+O_2$ | 5.3 | 4.3 | 4.3 | 3.6 | -3.8 | -5.5 | -4.9 | -16.8 | 0.39 |
| $M(O)(O_2)^S$ | $M(O)\cdot(O_2)^T$ | — | — | — | — | -18.8 | -19.8 | -19.6 | -21.8 | — |
| $M(O)(O_2)^S+E$ | $M(O)_2^S+EO$ | 22.1 | 22.7 | 22.2 | 34.3 | -5.7 | -6.6 | -5.7 | -8.0 | 0.51 |
| $M(O)_2^S$ | $M(O)_2^T$ | — | — | — | — | -7.0 | -6.1 | -5.5 | -8.1 | — |
| $M(O_2)_{2,spiro}^S$ | $M\cdot(O_2)_{2,spiro}^T$ | — | — | — | — | -17.4 | -18.1 | -17.4 | -19.0 | — |
| $M(O)\cdot(O_2)^T+O_2$ | $M(\eta^2\text{-}O_3)\cdot(O_2)^T$ | 16.9 | 19.3 | 17.9 | 32.5 | 8.5 | 12.6 | 11.4 | 25.3 | 0.50 |
| $M(\eta^2\text{-}O_3)\cdot(O_2)^T$ | $M(\eta^3\text{-}O_3)\cdot(O_2)^T$ | 4.7 | 2.3 | 2.1 | 2.2 | 3.8 | 3.5 | 2.4 | 6.7 | 0.58 |
| $M(\eta^3\text{-}O_3)\cdot(O_2)^T+E$ | $M\cdot(O_2)_{2,spiro}^T+EO$ | 12.1 | 11.2 | 12.2 | 19.2 | -23.5 | -23.9 | -23.1 | -28.4 | 0.45 |
| $M\cdot(O_2)_{2,spiro}^T+E$ | $M(O)\cdot(O_2)^T+EO$ | 21.5 | 23.2 | 22.8 | 35.3 | -16.4 | -14.8 | -15.1 | -16.1 | 0.57 |
| $M(O)\cdot(O_2)^T$ | $M(O)(O_2)^S$ | — | — | — | — | 18.8 | 19.8 | 19.6 | 21.8 | — |
| $M(O)(O_2)^S+O_2$ | $M(O_2)(\eta^2\text{-}O_3)^S$ | 17.4 | 19.0 | 17.8 | 30.2 | 0.7 | 2.7 | 1.9 | 13.6 | 0.43 |
| $M(O_2)(\eta^2\text{-}O_3)^S$ | $M(O_2)(\eta^3\text{-}O_3)^S$ | 5.3 | 5.0 | 4.6 | 4.5 | 4.3 | 4.5 | 4.3 | 5.3 | 0.54 |
| $M(O_2)(\eta^3\text{-}O_3)^S+E$ | $M(O_2)_{2,spiro}^S+EO$ | 14.0 | 14.8 | 14.6 | 25.9 | -17.6 | -16.8 | -17.6 | -18.1 | 0.44 |
| $M(O_2)_{2,spiro}^S+E$ | $M(O)(O_2)^S+EO$ | 21.5 | 22.8 | 21.6 | 32.7 | -15.0 | -13.1 | -13.0 | -13.3 | 0.59 |
| $M(O_2)_{2,spiro}^S$ | $M(O_2)_{2,planar}^S$ | 7.1 | 6.8 | 6.5 | 7.7 | -3.1 | -2.8 | -2.3 | -2.1 | 0.74 |
| $M(O_2)_{2,planar}^S+E$ | $M(O)(O_2)^S+EO$ | 22.9 | 24.2 | 23.7 | 35.8 | -11.8 | -10.3 | -10.7 | -11.2 | 0.59 |
| $M\cdot(O_2)_{2,butterfly}^T$ | $M(O_2)^T+O_2$ | 19.2 | 18.4 | 17.8 | 19.2 | -1.6 | -2.9 | -2.6 | -13.5 | 0.30 |
| $M(O_2)^T+E$ | $M(O)^T+EO$ | 20.8 | 22.5 | 21.7 | 36.7 | -12.1 | -10.3 | -11.4 | -9.0 | 0.66 |
| $M(O)^T+O_2$ | $M(\eta^2\text{-}O_3)^T$ | 21.4 | 17.2 | 18.5 | 25.1 | 6.7 | 9.0 | 8.5 | 20.1 | CP |
| $M(\eta^2\text{-}O_3)^T$ | $M(\eta^3\text{-}O_3)^T$ | 4.2 | 3.6 | 3.4 | 4.1 | 3.4 | 2.8 | 3.0 | 2.1 | 0.54 |
| $M(\eta^3\text{-}O_3)^T+E$ | $M(O_2)^T+EO$ | 15.5 | 16.7 | 15.6 | 30.6 | -25.5 | -24.0 | -24.5 | -25.7 | 0.58 |
| $M(\eta^2\text{-}O_3)^T$ | $M(\eta^2\text{-}O_3)^S$ | — | — | — | — | 11.2 | 11.4 | 11.4 | 11.8 | — |
| $M(\eta^2\text{-}O_3)^S$ | $M(\eta^3\text{-}O_3)^S$ | 4.6 | 4.1 | 3.8 | 4.5 | 3.8 | 3.6 | 3.6 | 4.3 | 0.63 |
| $M(\eta^3\text{-}O_3)^S+E$ | $M(O_2)^S+EO$ | 16.0 | 17.2 | 16.7 | 29.6 | -25.1 | -23.9 | -24.2 | -25.8 | 0.64 |
| $M(O_2)^S+ E$ | $M(O)^S+EO$ | 20.3 | 22.1 | 21.3 | 36.4 | -11.7 | -9.9 | -10.4 | -10.7 | 0.65 |
| $M(O)^S$ | $M(O)^T$ | — | — | — | — | -12.5 | -12.9 | -13.2 | -12.2 | — |

**Table 2** Energetic spans (kcal/mol) for direct ethylene epoxidation over four Zr/Hf-based catalysts using molecular oxygen as oxidant without coreductant. Energetic spans for Zr_NCCN, Zr_NCCNO, and Hf_NCCNO catalysts are from reference [22]

| **Catalyst** | $E_{SCF}$ | $E_{ZP}$ | **H** | **G** |
|---|---|---|---|---|
| Zr_NCCN | 53.1 | 52.6 | 52.7 | 62.0 |
| Zr_NCCNO | 30.2 | 30.5 | 30.2 | 44.7 |
| Hf_NCCNO | 25.8 | 26.1 | 25.0 | 45.3 |
| Zr_Benzol | 24.4 | 28.3 | 27.1 | 52.9 |

### 3.4 Quantifying Charge Transfer and Geometric Changes during Key Reaction Cycles

For the purpose of understanding the electronic structure of the Zr_Benzol catalyst and its charge transfer properties during an olefin epoxidation reaction, net atomic charges (NACs) and bond orders were evaluated using the Density-Derived Electrostatic and Chemical (DDEC6) method [35-37]. The computed NACs and bond orders for every atom in the various ground and transition state structures are given in Online Resource 2. We now summarize key results for intermediates in the rate-determining cycles (1st and 3rd junior cycles). The DDEC method has the conceptual advantage of representing NACs, ASMs, and bond orders as functionals of the electron and spin density distributions with no explicit basis set dependence. [35-37] The DDEC NACs are simultaneously optimized to reproduce the chemical states of atoms in a material and the electrostatic potential surrounding the material with excellent conformational transferability. [35-37] This makes the DDEC NACs, ASMs, and bond orders well-suited for studying charge transfer, spin transfer, and bond order changes during chemical reactions. Computed results are summarized in Table 3, Figs. 12, 13, and 14. For conciseness, the ligand atoms are not shown in Figs. 12–14, but the summed ligand atoms net charge and metal-ligand bond orders are summarized in Table 3. Figs. 12–14 also show selected bond lengths. For the situation where atoms or bonds are in a same (or very similar) chemical environment, only one of these atoms or bonds was labeled to avoid redundant information. For example, for a non-reacting peroxo group, only one O atom was labeled with charge and only one Zr-O bond along with the O-O bond were labeled with bond orders and bond lengths. Under such circumstance, all labeled values are the average values between the chemically equivalent atoms or bonds. For comparison, the computed charges, bond orders, and bond lengths for the $O_2$, ethylene, and ethylene oxide molecules are given in Fig. 14.

Table 3 summarizes the sum of bond orders (SBOs) for the Zr atom, the total bond orders between Zr and the two bidentate ligands, and the total charge residing on both ligands for each intermediate. (These values are not per ligand, they are the total values.) For every ground and transition state we tested, the Zr atom was positively charged with a charge of around 2.2 and the SBO for Zr is about 2.7. This stableness of the charge and SBO indicates that the Zr atom functions as an electron transfer bridge between the oxygenated group(s) and the bidentate redox ligands. As shown in Table 3, complexes with no strongly adsorbed oxygenated groups formed higher total metal-ligand bond orders than those with either one or two strongly adsorbed oxygenated groups. Similarly, complexes with one strongly adsorbed oxygenated group formed higher total metal-ligand bond orders than those with two strongly adsorbed oxygenated groups. This accounts for the almost constant SBO for the Zr atom. Complexes with a smaller total metal-ligands bond order also displayed less negative total ligand charges. This shows that increased bonding between the metal and the ligands also leads to more negatively charged ligands.

Bond orders, bond lengths, and NACs for the adsorbed oxygenated groups followed expected chemical trends. First, Zr-O bond orders are higher for strongly adsorbed O atoms than for weakly adsorbed O atoms. Also, the NACs for strongly adsorbed O atoms were more negative than those for weakly adsorbed O atoms. A total of approximately one electron was transferred to strongly adsorbed oxo, peroxo, or ozone groups. Second, for a given pair of elements, the bond orders typically increase as the bond length decreases. Third, the transfer of electrons to the adsorbed $O_2$ and $O_3$ groups was accompanied by longer O-O bond lengths and lower O-O bond orders. Because the low-lying unoccupied orbitals of $O_2$ and $O_3$ are $\pi^*$ (anti-bonding) orbitals, the transfer of electrons into the $O_2$ and $O_3$ adsorbate groups weakens their O-O bonds. Accordingly, structures with more negative O atoms exhibited greater O-O bond weakening. This weakening of the O-O bonds makes it easier for substrates to remove one of these O atoms. Fourth, the computed Zr SBO was consistent with observed electron transfer trends. If the Zr organometallic atom were neutral, one would expect its SBO to be ~4 due to the sharing of its four valence electrons with the ligands and adsorbate groups. The Zr SBO of ~2.7 indicates polarized covalent bonding in which some electron density has been transferred from the Zr atom to the ligands and adsorbate groups. This agrees with the positive NAC for Zr. Fifth, the interaction between Zr and the μ-O is quite weak even in an $\eta^3$-ozone group with the computed bond order of ~0.2.

In the ethylene oxide formation transition states, the C-C bond order (bond length) was lower (longer) than for free ethylene. Specifically, the C-C bond order was ~2.2 in free ethylene but ~1.7 in the epoxidation transition states ($TS_1$, $TS_4$, $TS_{12}$ and $TS_{14}$). The optimized C-C bond length in these transition states was ~0.03 Å longer than in free ethylene. In these transition states, one of the ethylene C atoms was much closer than the other ethylene C atom to the reacting O atom. Also, ~0.2 electrons were transferred from ethylene to the reacting O atom.

For the ozone transformation transition states ($TS_3$ and $TS_{13}$), no obvious charge transfer is observed in both cases and a small bond order of ~0.1 was achieved between the μ-O and Zr atoms. This indicates the $\eta^2$-ozone to $\eta^3$-ozone reaction is mainly a geometric rearrangement with little change in bond orders or NACs, except for the weak bond formed between the μ-O and Zr atoms

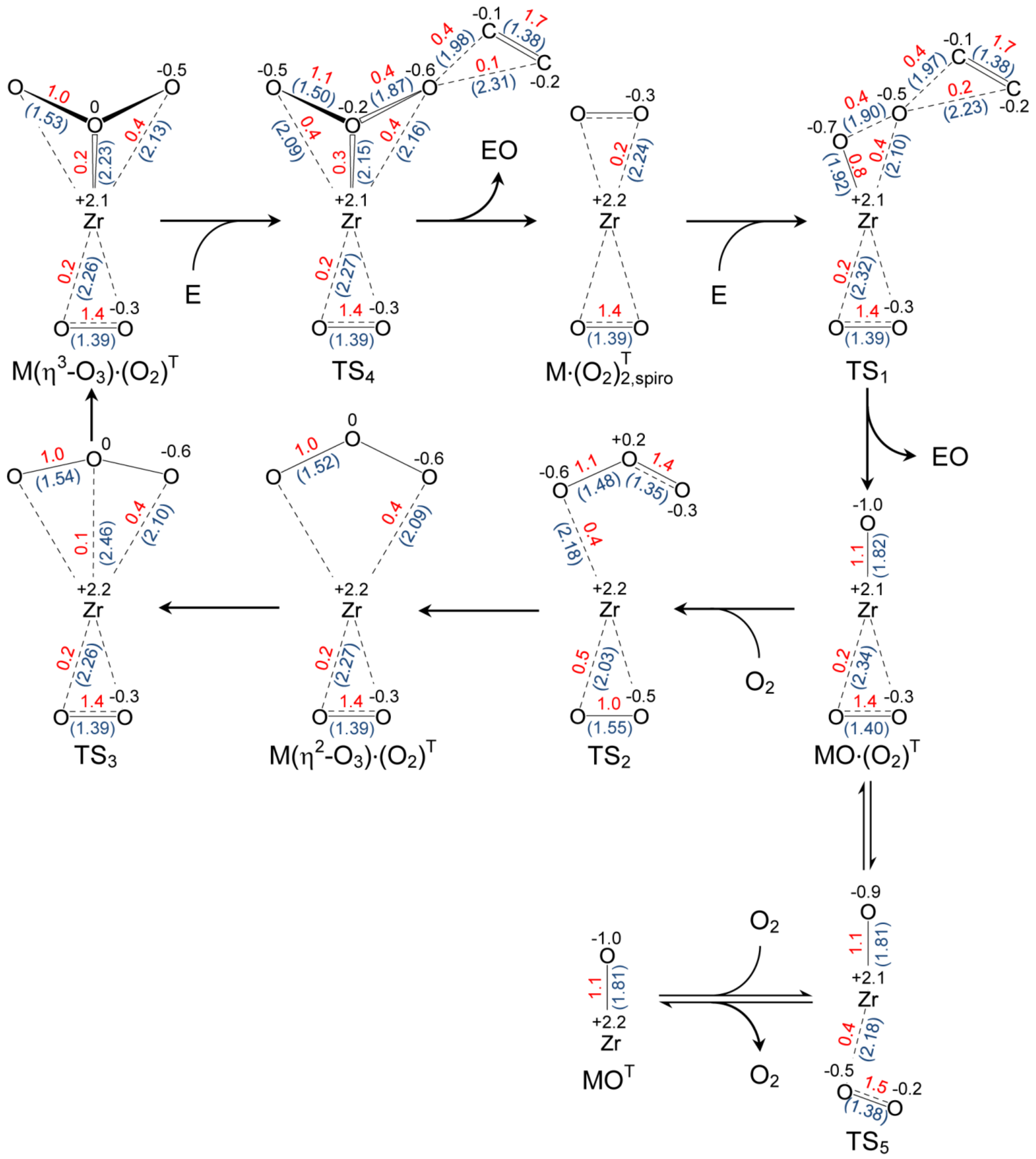

**Fig. 12** The 1st junior cycle for direct ethylene epoxidation. Atoms and chemical bonds in all intermediates are labeled with DDEC6 computed charges and bond orders, respectively. Bond lengths (Å) from the DFT-optimized geometries are given in parentheses. Ligands (and H atoms in ethylene) were included in all calculations but for conciseness are not shown here. Charges are presented in black with bond orders in red and bond lengths in blue

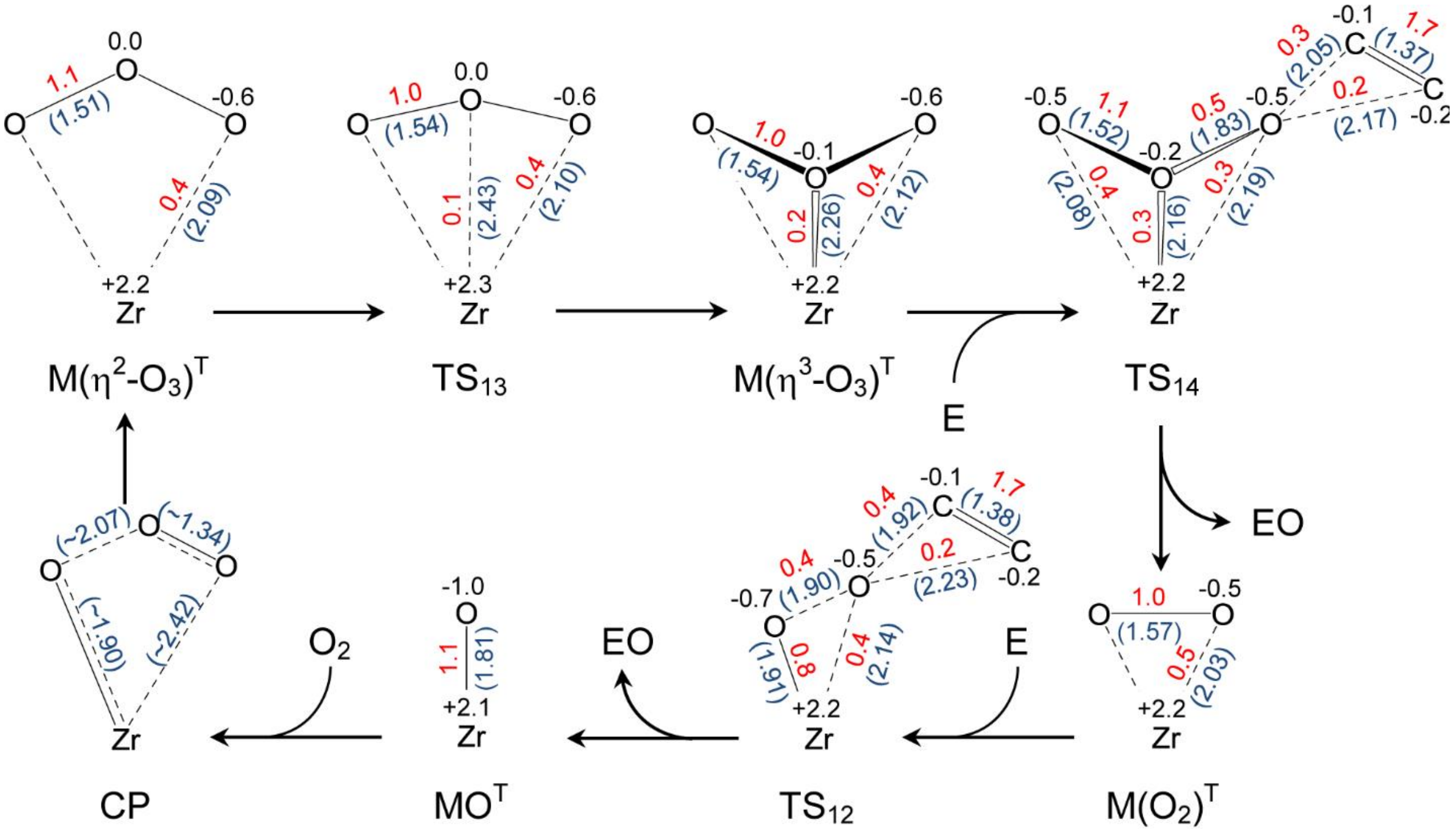

**Fig. 13** The 3rd junior cycle for direct ethylene epoxidation. Atoms and chemical bonds are labeled with DDEC6 computed charges and bond orders, respectively. Bond lengths (Å) from the DFT-optimized geometries are given in parentheses. Since the crossing point involves two different spin states, only its bond lengths are labeled. Ligands (and H atoms in ethylene) were included in all calculations but for conciseness are not shown here. Charges are presented in black with bond orders in red and bond lengths in blue

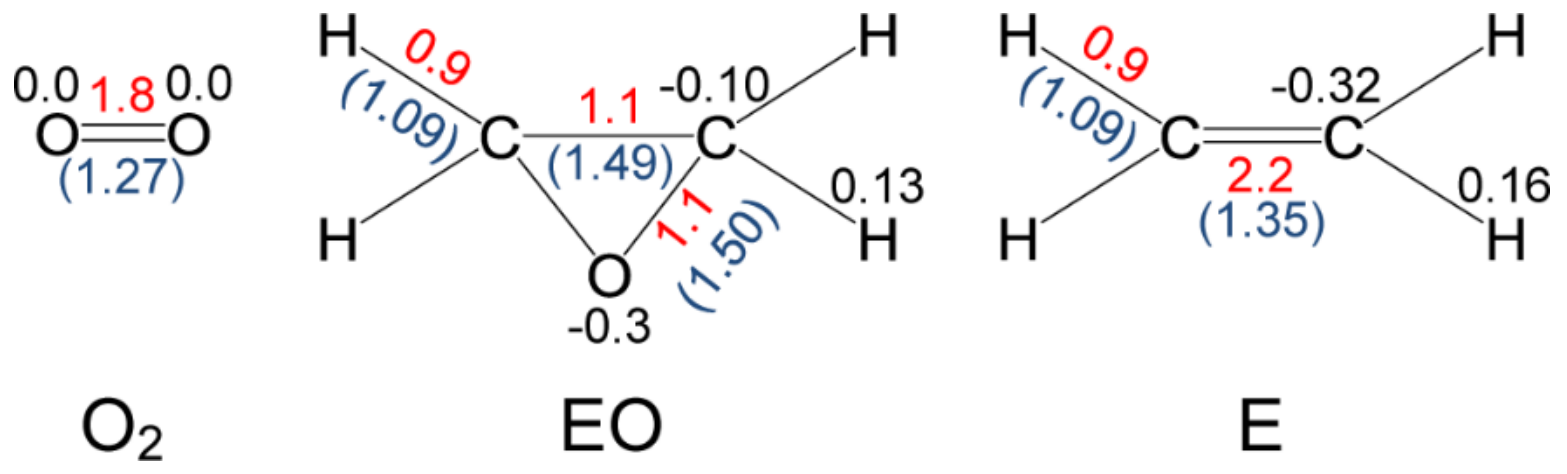

**Fig. 14** Summary of DDEC6 computed charges, bond orders, and bond lengths for the $O_2$, ethylene oxide, and ethylene molecules. Charges are presented in black with bond orders in red and bond lengths in blue parentheses

**Table 3** Summary of the DDEC6 results for the Zr sum of bond orders (Zr SBO), the total bond orders between Zr and two bidentate ligands (total ligands BOs), and the total charge residing on both ligands (total ligands charges) for each intermediate

| **Intermediates** | **Zr SBO** | **Total Ligands BOs** | **Total Ligands Charges** |
|---|---|---|---|
| $M(\eta^3\text{-}O_3)\cdot(O_2)^T$ | 2.6 | 1.2 | -0.5 |
| $TS_4$ | 2.6 | 1.1 | -0.5 |
| $M\cdot(O_2)^T{}_{2,spiro}$ | 2.7 | 1.9 | -1.0 |
| $TS_1$ | 2.7 | 1.1 | -0.6 |
| $MO\cdot(O_2)^T$ | 2.8 | 1.3 | -0.5 |
| $TS_2$ | 2.6 | 1.2 | -0.5 |
| $M(\eta^2\text{-}O_3)\cdot(O_2)^T$ | 2.6 | 1.4 | -0.4 |
| $TS_3$ | 2.6 | 1.3 | -0.4 |
| $TS_5$ | 2.8 | 1.3 | -0.5 |
| $M(\eta^2\text{-}O_3)^T$ | 2.6 | 1.8 | -1.0 |
| $TS_{13}$ | 2.6 | 1.7 | -1.1 |
| $M(\eta^3\text{-}O_3)^T$ | 2.6 | 1.6 | -1.0 |
| $TS_{14}$ | 2.6 | 1.6 | -1.3 |
| $M(O_2)^T$ | 2.7 | 1.7 | -1.2 |
| $TS_{12}$ | 2.7 | 1.5 | -1.3 |
| $MO^T$ | 2.9 | 1.8 | -1.1 |

### 3.5 Effects of Different Levels of Theory

**Table 4** Summary of computed cycle activation barriers (kcal/mol) for the rate determining cycles (1st & 3rd junior cycles) at different levels of theory. Overall energetic spans for direct ethylene epoxidation over the Zr_Benzol catalyst are highlighted in bold

| **Levels of Theory** | **SCF** | | **ZP** | | **H** | | **G** | |
|---|---|---|---|---|---|---|---|---|
| | 1st | 3rd | 1st | 3rd | 1st | 3rd | 1st | 3rd |
| B3LYP/LANL2DZ | **24.4** | 25.6 | **28.3** | 28.4 | 27.3 | **27.1** | 60.9 | **52.9** |
| B3LYP/LANL2DZ w/o loose | 25.7 | **25.6** | 30.6 | **28.4** | 29.4 | **27.7** | 63.7 | **51.0** |
| B3LYP/LANL2DZ w/ PCM | **26.0** | 26.8 | **29.7** | 29.9 | **28.2** | 28.4 | 63.2 | **55.3** |
| B3LYP/custom basis set | **30.4** | 33.7 | **34.5** | 36.5 | **32.9** | 35.1 | 67.8 | **60.8** |
| B3LYP/custom basis set w/ PCM | 33.0 | n/a[a] | 37.5 | n/a[a] | 36.3 | n/a[a] | 70.3 | n/a[a] |

[a] We were unable to converge this calculation in spite of trying many different convergence algorithms.

Different levels of theory were tested to justify the appropriateness of the choice of our computational model. For the sake of brevity and considering the large computational cost of calculations involving large basis sets, only intermediates and transition states that directly related to the overall energetic span of the catalytic cycle were re-optimized. Specifically, the $MO^T$ and $M(O)\cdot(O_2)^T$ ground state geometries and the $TS_4$ and $TS_{14}$ transition state geometries in the 1$^{st}$ or 3$^{rd}$ junior cycles were reoptimized for each level of tested theory. The overall energetic spans for direct ethylene epoxidation were recomputed accordingly.

Levels of theory we evaluated included: a) B3LYP/LANL2DZ using the "opt=loose" optimization convergence criteria (0.01 bohr for displacements, and 0.0025 a.u. for forces) and without any solvent, b) B3LYP/LANL2DZ using the tight optimization convergence criteria (0.0018 bohr for displacements, and 0.00045 a.u. for forces) and without any solvent, c) B3LYP/LANL2DZ using the "opt=loose" optimization convergence criteria combined with implicit toluene solvent using the Polarizable Continuum Model (PCM) [43-47], d) B3LYP paired with a custom basis set (LANL2DZ basis set for the H and ligand-related C atoms, LANL2DZdp basis set for the N and O atoms and the C atoms in ethylene/ethylene oxide, LANL2TZ(f) basis set for the Zr atom (which uses the same effective core potential as LANL2DZ) using the "opt=loose" optimization convergence criteria without any solvent, e) B3LYP paired with the custom basis set using the "opt=loose" optimization convergence criteria combined with implicit toluene solvent (PCM). The C, N, and O LANL2DZdp basis sets add a diffuse p function and a polarizing d function to the LANL2DZ basis set. The Zr LANL2TZ(f) basis set includes a polarizing f function, a triple zeta valence space, a diffuse s function, and a diffuse p function. The PCM calculations included the self-consistent reaction field (SCRF) energies that describe the electronic polarization due to solvent dielectric effects, but do not include the dispersion or cavitation energies. (The dispersion and cavitation energies were omitted because they are not computed self-consistently.)

Table 4 summarizes the computed cycle activation barriers for the rate determining cycles (1st & 3rd junior cycles) at different levels of theory. Overall energetic spans for direct ethylene epoxidation over the Zr_Benzol catalyst are highlighted in bold. SCF energies, zero-point energies, enthalpies, and Gibbs free energies are reported. The cycle activation barriers for the 1$^{st}$ and 3$^{rd}$ cycles are similar and no major differences were observed for the different levels of theory tested. For the enthalpic energetic span, the computed values ranged from 27.1 to 32.9 kcal/mol with an average value of 29.0 kcal/mol and a standard deviation of 2.7 kcal/mol. This demonstrates that the B3LYP/LANL2DZ level of theory is good enough for determining key trends in the performance of our Zr_Benzol catalyst for direct ethylene epoxidation.

**3.6 Propene Epoxidation**

All current industrial propylene oxide manufacturing processes generate a coproduct and require a coreductant or oxidant besides molecular $O_2$. [14, 20] In some cases, such as when using hydrogen peroxide as the oxidant, the only coproduct is water; however, this process still consumes one mole of $H_2$ gas per mole of PO produced in order to generate the $H_2O_2$ oxidant. [14, 16, 20, 48-55] Direct propene epoxidation that uses only molecular oxygen as oxidant without a coreductant has long been desired, because it would eliminate coreactants and coproducts. Large amount of efforts have been made in new catalyst design at laboratory level for direct propene epoxidation using molecular oxygen as oxidant without coreductant, but none were commercialized yet. [11, 14, 56-63]

We used DFT computations to extensively study the mechanism of direct propene epoxidation over the Zr_Benzol catalyst using molecular oxygen as oxidant without coreductant. The computed catalytic cycles, activation barriers, and energy profiles for direct propene epoxidation are similar to those for direct ethylene epoxidation. (Detailed information including reaction cycles and energy tables are in the Online Resource 1.) However, our computations for direct propene epoxidation also revealed an undesirable side reaction that is not present for direct ethylene epoxidation. Fig. 15 shows the deactivation product for the Zr_Benzol catalyst that can form when the alkene, such as propene, contains an allylic hydrogen atom. Since ethylene has no allylic hydrogen atoms, it cannot form this type of deactivation product. In this deactivation product, the allylic hydrogen atom is transferred from the alkene or alkene oxide to form a Zr–O–H group and breaks the C–H(allylic) bond. As shown Fig. 15, the reactant contains a Zr–O–$CH_2$–CH–$CH_3$ chain plus an oxo group bound to the Zr atom. Table 5 summarizes the energies of the allylic H transfer reactant and product relative to the triplet oxo form of the catalyst plus PO. The low relative energies (-27.3 to -38.6 kcal/mol) of the allylic H transfer product compared to forming the desired PO product show this unwanted side reaction is thermodynamically preferred. In Table 5, the transition state energies are referenced to the reactant state and equal the activation barriers for the allylic H transfer reaction. The low activation barriers (5.1 to 7.8 kcal/mol) indicate this unwanted side reaction would occur readily under reaction conditions. Unfortunately, these results indicate the Zr_Benzol catalyst is not efficient for direct propene epoxidation and probably will not perform well for other alkenes that have allylic hydrogen atoms.

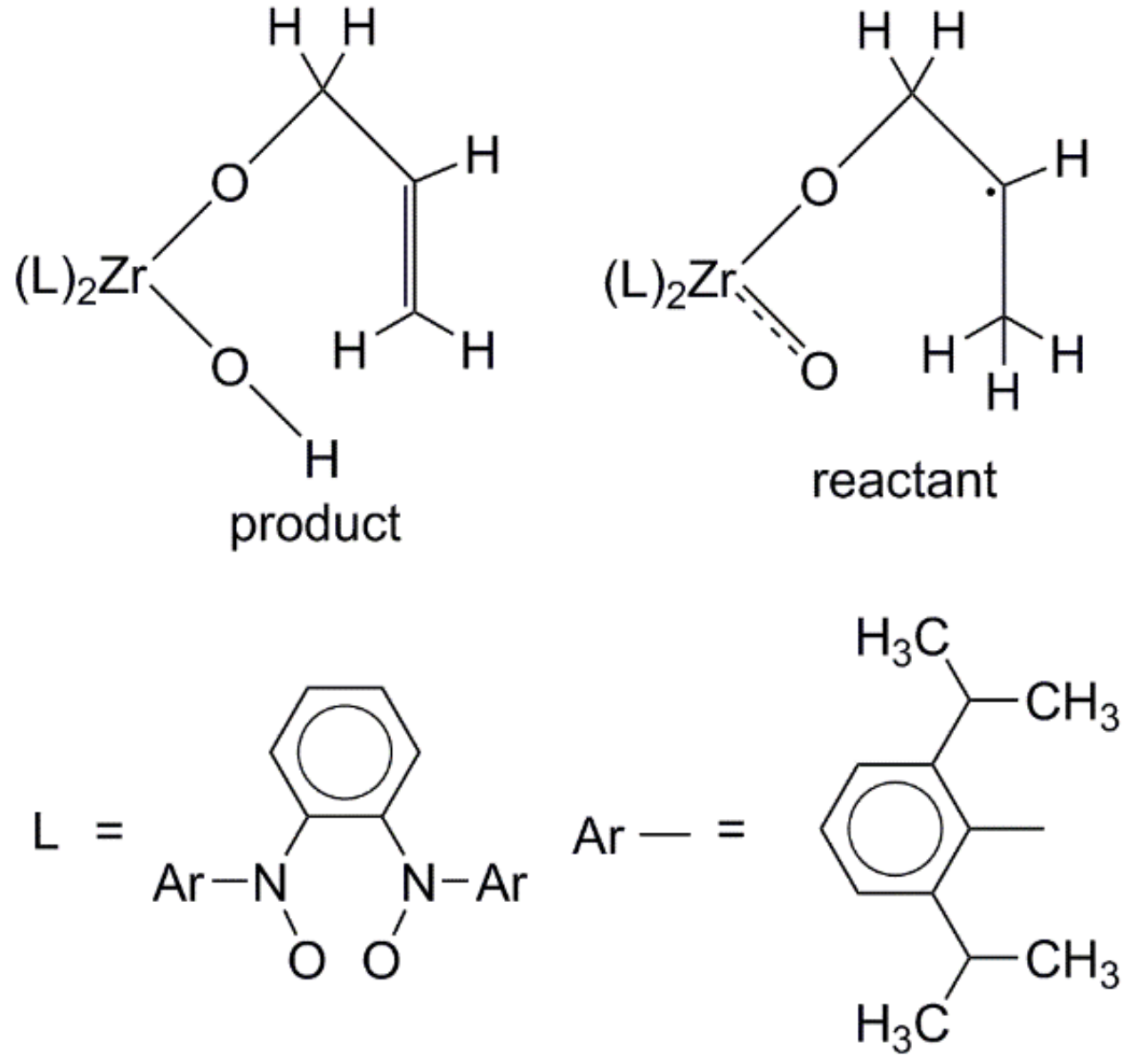

**Fig. 15** Undesirable by-product that forms from the Zr_Benzol catalyst when using propene reactant. Left: Product state with an allylic hydrogen atom transferred to catalyst. Right: Reactant state

Table 6 summarizes the overall energetic barriers for catalyzed ethylene and propene epoxidation computed by other research groups. This may not be a complete list of all previously computed energetic spans for ethylene and propene epoxidation, but it should contain enough examples to be representative. Lundin et al. used the B3LYP/6-311+G(d,p) level of DFT simulations to study the mechanism of propene and ethylene epoxidation using hydrogen peroxide as the oxidant. [64] According to the reaction energy diagrams they provided, enthalpic energetic spans of 17 (ethylene epoxidation) and 13 (propene epoxidation) kcal/mol were achieved for a binuclear Ti dihydroxide catalyst. [64] de Visser et al. used the B3LYP/6-311++G** level of DFT simulations to study the mechanism of ethylene and propene epoxidation using hydrogen peroxide as the oxidant in the presence of various fluorinated alcohols in methanol solvent. [65] According to the reaction energy diagrams they provided, electronic energetic spans of 25 to 35 (ethylene epoxidation) and 21 to 31 (propene epoxidation) kcal/mol were achieved. [65] Joshi et al. studied the “sequential” and “simultaneous” mechanism of propene epoxidation using $H_2O_2$ over Au/titanosilicate-1 catalysts. [66] According to their QM/MM calculations, an electronic energetic span of 20 kcal/mol is achieved on the Ti-defect site. [66] Lei et al. studied direct propene oxidation using molecular oxygen as the oxidant over subnanometer silver particles. [11] An enthalpic energetic span around 41 kcal/mol was achieved. [11] Comas-Vivas et al. used the B3LYP exchange-correlation functional to study ethylene epoxidation using hydrogen peroxide and MeOOH oxidants over a cyclopentadienyl-Mo complex. [67] According to the reaction energy diagrams they provided, electronic energetic spans of 27 ($H_2O_2$ oxidant) and 27 (MeOOH oxidant) kcal/mol were achieved. [67] Dinoi et al. used the B3PW91/SDD level of DFT simulations to study ethylene epoxidation using hydrogen peroxide as oxidant over a cyclopentadienyloxidotungsten complex. [68] According to the reaction energy diagrams they provided, an enthalpic energetic span of 27 kcal/mol is achieved. [68] Herbert et al. used the B3LYP exchange-correlation functional to study ethylene epoxidation using hydrogen peroxide as oxidant over an oxodiperoxomoly complex with a computed free energy energetic span of ~32 kcal/mol. [69] Kuznetsov et al. used the B3LYP exchange-correlation functional to study ethylene epoxidation using hydrogen peroxide as oxidant over a vanadium-salan model complex with a computed enthalpic energetic span of ~33 kcal/mol. [70]

**Table 5** Summary of the computed energies of the reactant, transition state, and product for the allylic hydrogen transfer reaction. The reference state for reactants and products is the triplet oxo complex plus a propylene oxide molecule. The transition state energy is referenced to the reactant state and equal the activation barrier for the allylic H transfer reaction. Similar results for the Zr_NCCNO, Hf_NCCNO, Zr_ONCCNO, and Hf_ONCCNO catalysts were reported in references [21] and [22]

| **Catalyst** | **Structure** | **$E_{SCF}$ (kcal/mol)** | **$E_{ZP}$ (kcal/mol)** | **H (kcal/mol)** | **G (kcal/mol)** |
|---|---|---|---|---|---|
| Zr_Benzol | reactant | -2.3 | -1.1 | -0.3 | 10.0 |
| Zr_Benzol | transition state | 7.8 | 5.7 | 5.1 | 7.4 |
| Zr_Benzol | product | -38.6 | -38.2 | -37.2 | -27.3 |

**Table 6** Reference energetic spans for propene and ethylene epoxidation processes using different catalysts and oxidants

| Olefin | Oxidant | Catalyst | E span (kcal/mol) |
|---|---|---|---|
| propene | $O_2$ | subnanometer silver particles on alumina support [11] | ~41[a] |
| propene | $H_2O_2$ | binuclear Ti dihydroxide site [64] | 13[a] |
| propene | $H_2O_2$ | fluorinated alcohols in methanol [65] | 21 to 31[b] |
| propene | $H_2O_2$ | titanosilicate-1, Ti-defect [66] | 20[b] |
| ethylene | $H_2O_2$ | fluorinated alcohols in methanol [65] | 25 to 35[b] |
| ethylene | $H_2O_2$ | binuclear Ti dihydroxide site [64] | 17[a] |
| ethylene | $H_2O_2$ | cyclopentadienyl-Mo complex [67, 71] | 27[b] |
| ethylene | MeOOH | cyclopentadienyl-Mo complex [67, 71] | 27[b] |
| ethylene | $H_2O_2$ | cyclopentadienyloxidotungsten complex [68] | 27[a] |
| ethylene | $H_2O_2$ | oxodiperoxomoly complex [69] | ~32[c] |
| ethylene | $H_2O_2$ | vanadium-salan model complex [70] | ~33[a] |

[a] Enthalpy; [b] electronic energy; [c] Gibbs free energy

## 4 Proposed Synthesis Reaction

Here we propose a potential synthesis reaction for the Zr_Benzol catalyst. Here we choose tetrakis(dimethylamido)zirconium (i.e., $[(CH_3)_2N]_4Zr$), which can be purchased through chemical supply companies, as an example Zr metal provider to demonstrate the catalyst synthesis reaction. Similar reagents such as tetrakis(diethylamido)zirconium (i.e., $[(CH_3CH_2)_2N]_4Zr$) and tetrakis(ethylmethylamido)zirconium (i.e., $[(CH_3CH_2)(CH_3)N]_4Zr$) can also be purchased through chemical supply companies and should work similarly as Zr metal providers. As shown in Fig. 16, one mole of tetrakis(dimethylamido)zirconium could react with two moles of di(hydroxylamine) ligand to produce one mole of Zr_Benzol bare complex plus 4 moles of dimethylamine as products. The computed net reaction energies associated with this reaction are -65 ($E_{SCF}$), -62.4 ($E_{ZP}$), -62.6 (H), and -76.4 (G) kcal/mol. These indicate the catalyst synthesis reaction should be exothermic and thermodynamically favorable. For comparison, computed net reaction energies for the unwanted reaction $Zr(N(Me)_2)_4$ + 2 $C_6H_4$-1,6-($N(C_6H_3$-2',6'-$(CH(CH_3)_2)_2)OH)_2$ → $Zr(C_6H_4$-1,6-$(N(C_6H_3$-2',6'-$(CH(CH_3)_2)_2))_2)_2$ + 4 $N(Me)_2OH$ are 3.6 ($E_{SCF}$), 5.0 ($E_{ZP}$), 4.3 (H), and -9.4 (G) kcal/mol. These energies clearly indicate the oxygen atoms will be retained on the ligand to produce the desired Zr_Benzol catalyst plus dimethylamine instead of being extracted to produce a bis(diimine) Zr complex plus dimethylhydroxylamine.

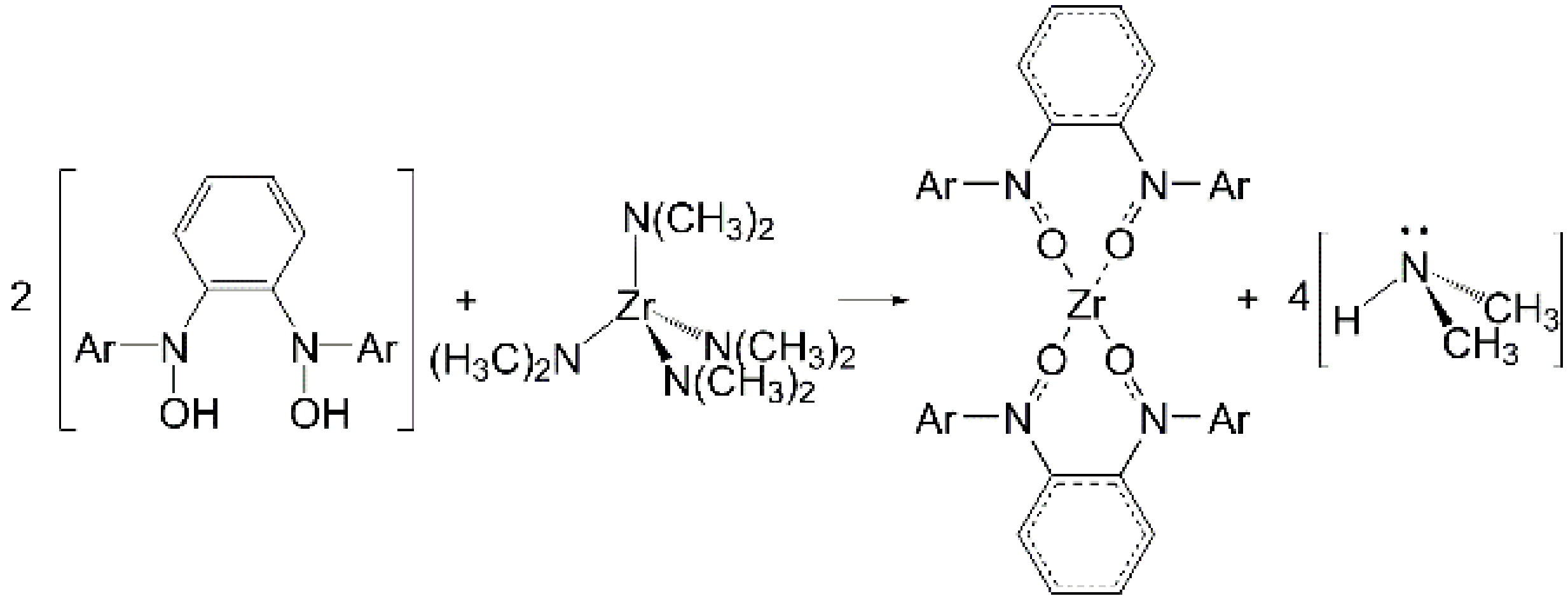

**Fig. 16** Proposed synthesis reaction for the Zr_Benzol catalyst. Reactants: Bare di(hydroxylamine) ligand and tetrakis(dimethylamido)zirconium. Products: Zr_Benzol bare complex and dimethylamine

## 5 Conclusions

In this article, we introduced a new Zr-based organometallic complex (Zr_Benzol) that can selectively oxidize organic substrates (e.g., ethylene) using molecular oxygen as the oxidant without a coreductant. This catalyst could potentially be synthesized via the condensation reaction $Zr(N(R)R')_4$ + 2 $C_6H_4$-1,6-($N(C_6H_3$-2',6'-($CH(CH_3)_2)_2)OH)_2$ → $Zr(C_6H_4$-1,6-($N(C_6H_3$-2',6'-($CH(CH_3)_2)_2)O)_2)_2$ [aka Zr_Benzol ] + 4 N(R)(R')H where R and R' are $CH_3$, $CH_2CH_3$, or other alkyl groups. We used DFT computations to study reaction mechanisms, energy profiles, and possible side reactions for ethylene epoxidation. No coreductant is required and no coproduct will be produced under idealized conditions. Our calculations revealed four interrelated junior cycles. Each junior cycle consumes one $O_2$ molecule and two ethylene molecules to produce two ethylene oxide molecules. Each catalytic cycle involves four key steps: a) addition of an $O_2$ molecule to an oxo group to generate an $\eta^2$-ozone group, b) rearrangement of the $\eta^2$-ozone group to form an $\eta^3$-ozone group, c) reacting the $\eta^3$-ozone group with a substrate (e.g., ethylene) molecule to produce a substrate oxide (e.g., ethylene oxide) molecule plus a peroxo or weakly adsorbed $O_2$ group, and d) reacting the peroxo or weakly adsorbed $O_2$ group with a substrate (e.g., ethylene) molecule to produce a substrate oxide (e.g., ethylene oxide) molecule and regenerate the oxo group. For direct ethylene epoxidation, a computed enthalpic energetic span (i.e., effective activation energy for the entire catalytic cycle) of 27.1 kcal/mol is achieved, which is one of the lowest values for catalysts studied to date. Selected computations at different levels of theory, including implicit solvation and larger basis sets, produced similar overall energetic spans.

Computations were performed to identify potential side reactions and deactivation products. For direct ethylene epoxidation, no energetically important deactivation products or side reactions were identified. Notably, this catalyst does not deactivate to form octahedral-like structures. Complete catalytic cycles and reaction energy profiles were also computed for direct propene epoxidation. Unfortunately, a side reaction that transfers an allylic hydrogen atom from alkene to catalyst makes this catalyst unsuitable for epoxidizing alkenes such as propene that contain allylic hydrogen atoms. Specifically, the unwanted allylic hydrogen transfer reaction makes the computed enthalpic energetic span for direct propene epoxidation so high (63.6 kcal/mol) the catalyst becomes ineffective. We recommend future computational screening studies to identify modifications that may be able to prevent the unwanted transfer of allylic hydrogen atoms.

Net atomic charges (NACs), bond orders, and transition state lateness descriptors were computed to gain chemical insights into the reaction chemistry. These computations showed the NAC and sum of bond orders for the Zr atom is almost unchanged during the key catalytic cycles. Therefore, the main electron transfer effects occur between the redox active bidentate ligands, the adsorbed oxygenated groups, and the reacting substrate molecule. In particular, the redox active bidentate ligands act as an electron bank, allowing electrons to be deposited or withdrawn as needed by the adsorbing and reacting groups. In turn, this allows the catalyst to operate as an efficient oxygen sponge that allows substrate molecules to deposit or extract O atoms with little change in the O atom chemical potential.

**Conflict of Interest**: The authors and NMSU's Office of Intellectual Property (Arrowhead Center, Inc.) have applied for a patent on some of the results described in this paper.